\ifx\mnmacrosloaded\undefined 
%
%
%
%

\catcode `\@=11 

\def\@version{1.6}
\def\@verdate{18th September 1995}

%
%


\newif\ifprod@font

\ifx\@typeface\undefined
  \def\@typeface{Comp. Modern}\prod@fontfalse
\else
  \prod@fonttrue 
\fi

\def\newfam{\alloc@8\fam\chardef\sixt@@n} 

\ifprod@font
\font\fiverm=mtr10 at 5pt
\font\fivebf=mtbx10 at 5pt
\font\fiveit=mtti10 at 5pt
\font\fivesl=mtsl10 at 5pt
\font\fivett=cmtt8 at 5pt     \hyphenchar\fivett=-1
\font\fivecsc=mtcsc10 at 5pt
\font\fivesf=mtss10 at 5pt
\font\fivei=mtmi10 at 5pt      \skewchar\fivei='177
\font\fivesy=mtsy10 at 5pt     \skewchar\fivesy='60

\font\sixrm=mtr10 at 6pt
\font\sixbf=mtbx10 at 6pt
\font\sixit=mtti10 at 6pt
\font\sixsl=mtsl10 at 6pt
\font\sixtt=cmtt8 at 6pt      \hyphenchar\sixtt=-1
\font\sixcsc=mtcsc10 at 6pt
\font\sixsf=mtss10 at 6pt
\font\sixi=mtmi10 at 6pt       \skewchar\sixi='177
\font\sixsy=mtsy10 at 6pt      \skewchar\sixsy='60

\font\sevenrm=mtr10 at 7pt
\font\sevenbf=mtbx10 at 7pt
\font\sevenit=mtti10 at 7pt
\font\sevensl=mtsl10 at 7pt
\font\seventt=cmtt8 at 7pt     \hyphenchar\seventt=-1
\font\sevencsc=mtcsc10 at 7pt
\font\sevensf=mtss10 at 7pt
\font\seveni=mtmi10 at 7pt      \skewchar\seveni='177
\font\sevensy=mtsy10 at 7pt     \skewchar\sevensy='60

\font\eightrm=mtr10 at 8pt
\font\eightbf=mtbx10 at 8pt
\font\eightit=mtti10 at 8pt
\font\eighti=mtmi10 at 8pt      \skewchar\eighti='177
\font\eightsy=mtsy10 at 8pt     \skewchar\eightsy='60
\font\eightsl=mtsl10 at 8pt
\font\eighttt=cmtt8             \hyphenchar\eighttt=-1
\font\eightcsc=mtcsc10 at 8pt
\font\eightsf=mtss10 at 8pt

\font\ninerm=mtr10 at 9pt
\font\ninebf=mtbx10 at 9pt
\font\nineit=mtti10 at 9pt
\font\ninei=mtmi10 at 9pt      \skewchar\ninei='177
\font\ninesy=mtsy10 at 9pt     \skewchar\ninesy='60
\font\ninesl=mtsl10 at 9pt
\font\ninett=cmtt9             \hyphenchar\ninett=-1
\font\ninecsc=mtcsc10 at 9pt
\font\ninesf=mtss10 at 9pt

\font\tenrm=mtr10
\font\tenbf=mtbx10
\font\tenit=mtti10
\font\teni=mtmi10		\skewchar\teni='177
\font\tensy=mtsy10		\skewchar\tensy='60
\font\tenex=cmex10
\font\tensl=mtsl10
\font\tentt=cmtt10		\hyphenchar\tentt=-1
\font\tencsc=mtcsc10
\font\tensf=mtss10

\font\elevenrm=mtr10 at 11pt
\font\elevenbf=mtbx10 at 11pt
\font\elevenit=mtti10 at 11pt
\font\eleveni=mtmi10 at 11pt      \skewchar\eleveni='177
\font\elevensy=mtsy10 at 11pt     \skewchar\elevensy='60
\font\elevensl=mtsl10 at 11pt
\font\eleventt=cmtt10 at 11pt     \hyphenchar\eleventt=-1
\font\elevencsc=mtcsc10 at 11pt
\font\elevensf=mtss10 at 11pt

\font\twelverm=mtr10 at 12pt
\font\twelvebf=mtbx10 at 12pt
\font\twelveit=mtti10 at 12pt
\font\twelvesl=mtsl10 at 12pt
\font\twelvett=cmtt12             \hyphenchar\twelvett=-1
\font\twelvecsc=mtcsc10 at 12pt
\font\twelvesf=mtss10 at 12pt
\font\twelvei=mtmi10 at 12pt      \skewchar\twelvei='177
\font\twelvesy=mtsy10 at 12pt     \skewchar\twelvesy='60

\font\fourteenrm=mtr10 at 14pt
\font\fourteenbf=mtbx10 at 14pt
\font\fourteenit=mtti10 at 14pt
\font\fourteeni=mtmi10 at 14pt      \skewchar\fourteeni='177
\font\fourteensy=mtsy10 at 14pt     \skewchar\fourteensy='60
\font\fourteensl=mtsl10 at 14pt
\font\fourteentt=cmtt12 at 14pt     \hyphenchar\fourteentt=-1
\font\fourteencsc=mtcsc10 at 14pt
\font\fourteensf=mtss10 at 14pt

\font\seventeenrm=mtr10 at 17pt
\font\seventeenbf=mtbx10 at 17pt
\font\seventeenit=mtti10 at 17pt
\font\seventeeni=mtmi10 at 17pt      \skewchar\seventeeni='177
\font\seventeensy=mtsy10 at 17pt     \skewchar\seventeensy='60
\font\seventeensl=mtsl10 at 17pt
\font\seventeentt=cmtt12 at 17pt     \hyphenchar\seventeentt=-1
\font\seventeencsc=mtcsc10 at 17pt
\font\seventeensf=mtss10 at 17pt
\else
\font\fiverm=cmr5
\font\fivei=cmmi5             \skewchar\fivei='177
\font\fivesy=cmsy5            \skewchar\fivesy='60
\font\fivebf=cmbx5

\font\sixrm=cmr6
\font\sixi=cmmi6             \skewchar\sixi='177
\font\sixsy=cmsy6            \skewchar\sixsy='60
\font\sixbf=cmbx6

\font\sevenrm=cmr7
\font\sevenit=cmti7
\font\seveni=cmmi7             \skewchar\seveni='177
\font\sevensy=cmsy7            \skewchar\sevensy='60
\font\sevenbf=cmbx7

\font\eightrm=cmr8
\font\eightbf=cmbx8
\font\eightit=cmti8
\font\eighti=cmmi8			\skewchar\eighti='177
\font\eightsy=cmsy8			\skewchar\eightsy='60
\font\eightsl=cmsl8
\font\eighttt=cmtt8			\hyphenchar\eighttt=-1
\font\eightcsc=cmcsc10 at 8pt
\font\eightsf=cmss8

\font\ninerm=cmr9
\font\ninebf=cmbx9
\font\nineit=cmti9
\font\ninei=cmmi9			\skewchar\ninei='177
\font\ninesy=cmsy9			\skewchar\ninesy='60
\font\ninesl=cmsl9
\font\ninett=cmtt9			\hyphenchar\ninett=-1
\font\ninecsc=cmcsc10 at 9pt
\font\ninesf=cmss9

\font\tenrm=cmr10
\font\tenbf=cmbx10
\font\tenit=cmti10
\font\teni=cmmi10		\skewchar\teni='177
\font\tensy=cmsy10		\skewchar\tensy='60
\font\tenex=cmex10
\font\tensl=cmsl10
\font\tentt=cmtt10		\hyphenchar\tentt=-1
\font\tencsc=cmcsc10
\font\tensf=cmss10

\font\elevenrm=cmr10 scaled \magstephalf
\font\elevenbf=cmbx10 scaled \magstephalf
\font\elevenit=cmti10 scaled \magstephalf
\font\eleveni=cmmi10 scaled \magstephalf	\skewchar\eleveni='177
\font\elevensy=cmsy10 scaled \magstephalf	\skewchar\elevensy='60
\font\elevensl=cmsl10 scaled \magstephalf
\font\eleventt=cmtt10 scaled \magstephalf	\hyphenchar\eleventt=-1
\font\elevencsc=cmcsc10 scaled \magstephalf
\font\elevensf=cmss10 scaled \magstephalf

\font\twelverm=cmr10 scaled \magstep1
\font\twelvebf=cmbx10 scaled \magstep1
\font\twelvei=cmmi10 scaled \magstep1      \skewchar\twelvei='177
\font\twelvesy=cmsy10 scaled \magstep1     \skewchar\twelvesy='60

\font\fourteenrm=cmr10 scaled \magstep2
\font\fourteenbf=cmbx10 scaled \magstep2
\font\fourteenit=cmti10 scaled \magstep2
\font\fourteeni=cmmi10 scaled \magstep2		\skewchar\fourteeni='177
\font\fourteensy=cmsy10 scaled \magstep2	\skewchar\fourteensy='60
\font\fourteensl=cmsl10 scaled \magstep2
\font\fourteentt=cmtt10 scaled \magstep2	\hyphenchar\fourteentt=-1
\font\fourteencsc=cmcsc10 scaled \magstep2
\font\fourteensf=cmss10 scaled \magstep2

\font\seventeenrm=cmr10 scaled \magstep3
\font\seventeenbf=cmbx10 scaled \magstep3
\font\seventeenit=cmti10 scaled \magstep3
\font\seventeeni=cmmi10 scaled \magstep3	\skewchar\seventeeni='177
\font\seventeensy=cmsy10 scaled \magstep3	\skewchar\seventeensy='60
\font\seventeensl=cmsl10 scaled \magstep3
\font\seventeentt=cmtt10 scaled \magstep3	\hyphenchar\seventeentt=-1
\font\seventeencsc=cmcsc10 scaled \magstep3
\font\seventeensf=cmss10 scaled \magstep3
\fi

\def\hexnumber#1{\ifcase#1 0\or1\or2\or3\or4\or5\or6\or7\or8\or9\or
  A\or B\or C\or D\or E\or F\fi}

\def\makestrut{%
  \setbox\strutbox=\hbox{%
    \vrule height.7\baselineskip depth.3\baselineskip width \z@}%
}

\def\baselinestretch{1}
\newskip\tmp@bls

\def\b@ls#1{
  \tmp@bls=#1\relax
  \baselineskip=#1\relax\makestrut
  \normalbaselineskip=\baselinestretch\tmp@bls
  \normalbaselines
}

\def\nostb@ls#1{
  \normalbaselineskip=#1\relax
  \normalbaselines
  \makestrut
}

%

\newfam\scfam  
\newfam\sffam  

\def\mit{\fam\@ne}
\def\cal{\fam\tw@}
\def\em{\ifdim\fontdimen1\font>\z@ \rm\else\it\fi}

\textfont3=\tenex
\scriptfont3=\tenex
\scriptscriptfont3=\tenex

\setbox0=\hbox{\tenex B} \p@renwd=\wd0 

\def\eightpoint{
  \def\rm{\fam0\eightrm}%
  \textfont0=\eightrm \scriptfont0=\sixrm \scriptscriptfont0=\fiverm%
  \textfont1=\eighti  \scriptfont1=\sixi  \scriptscriptfont1=\fivei%
  \textfont2=\eightsy \scriptfont2=\sixsy \scriptscriptfont2=\fivesy%
  \textfont\itfam=\eightit\def\it{\fam\itfam\eightit}%
  \ifprod@font
    \scriptfont\itfam=\sixit
      \scriptscriptfont\itfam=\fiveit
  \else
    \scriptfont\itfam=\eightit
      \scriptscriptfont\itfam=\eightit
  \fi
  \textfont\bffam=\eightbf%
    \scriptfont\bffam=\sixbf%
      \scriptscriptfont\bffam=\fivebf%
  \def\bf{\fam\bffam\eightbf}%
  \textfont\slfam=\eightsl\def\sl{\fam\slfam\eightsl}%
  \ifprod@font
    \scriptfont\slfam=\sixsl
      \scriptscriptfont\slfam=\fivesl
  \else
    \scriptfont\slfam=\eightsl
      \scriptscriptfont\slfam=\eightsl
  \fi
  \textfont\ttfam=\eighttt\def\tt{\fam\ttfam\eighttt}%
  \ifprod@font
    \scriptfont\ttfam=\sixtt
      \scriptscriptfont\ttfam=\fivett
  \else
    \scriptfont\ttfam=\eighttt
      \scriptscriptfont\ttfam=\eighttt
  \fi
  \textfont\scfam=\eightcsc\def\sc{\fam\scfam\eightcsc}%
  \ifprod@font
    \scriptfont\scfam=\sixcsc
      \scriptscriptfont\scfam=\fivecsc
  \else
    \scriptfont\scfam=\eightcsc
      \scriptscriptfont\scfam=\eightcsc
  \fi
  \textfont\sffam=\eightsf\def\sf{\fam\sffam\eightsf}%
  \ifprod@font
    \scriptfont\sffam=\sixsf
      \scriptscriptfont\sffam=\fivesf
  \else
    \scriptfont\sffam=\eightsf
      \scriptscriptfont\sffam=\eightsf
  \fi
  \def\oldstyle{\fam\@ne\eighti}%
  \b@ls{10pt}\rm\@viiipt%
}
\def\@viiipt{}

\def\ninepoint{
  \def\rm{\fam0\ninerm}%
  \textfont0=\ninerm \scriptfont0=\sixrm \scriptscriptfont0=\fiverm%
  \textfont1=\ninei  \scriptfont1=\sixi  \scriptscriptfont1=\fivei%
  \textfont2=\ninesy \scriptfont2=\sixsy \scriptscriptfont2=\fivesy%
  \textfont\itfam=\nineit\def\it{\fam\itfam\nineit}%
  \ifprod@font
    \scriptfont\itfam=\sixit
      \scriptscriptfont\itfam=\fiveit
  \else
    \scriptfont\itfam=\nineit
      \scriptscriptfont\itfam=\nineit
  \fi
  \textfont\bffam=\ninebf%
    \scriptfont\bffam=\sixbf%
      \scriptscriptfont\bffam=\fivebf%
  \def\bf{\fam\bffam\ninebf}%
  \textfont\slfam=\ninesl\def\sl{\fam\slfam\ninesl}%
  \ifprod@font
    \scriptfont\slfam=\sixsl
      \scriptscriptfont\slfam=\fivesl
  \else
    \scriptfont\slfam=\ninesl
      \scriptscriptfont\slfam=\ninesl
  \fi
  \textfont\ttfam=\ninett\def\tt{\fam\ttfam\ninett}%
  \ifprod@font
    \scriptfont\ttfam=\sixtt
      \scriptscriptfont\ttfam=\fivett
  \else
    \scriptfont\ttfam=\ninett
      \scriptscriptfont\ttfam=\ninett
  \fi
  \textfont\scfam=\ninecsc\def\sc{\fam\scfam\ninecsc}%
  \ifprod@font
    \scriptfont\scfam=\sixcsc
      \scriptscriptfont\scfam=\fivecsc
  \else
    \scriptfont\scfam=\ninecsc
      \scriptscriptfont\scfam=\ninecsc
  \fi
  \textfont\sffam=\ninesf\def\sf{\fam\sffam\ninesf}%
  \ifprod@font
    \scriptfont\sffam=\sixsf
      \scriptscriptfont\sffam=\fivesf
  \else
    \scriptfont\sffam=\ninesf
      \scriptscriptfont\sffam=\ninesf
  \fi
  \def\oldstyle{\fam\@ne\ninei}%
  \b@ls{\TextLeading plus \Feathering}\rm\@ixpt%
}
\def\@ixpt{}

\def\tenpoint{
  \def\rm{\fam0\tenrm}%
  \textfont0=\tenrm \scriptfont0=\sevenrm \scriptscriptfont0=\fiverm%
  \textfont1=\teni  \scriptfont1=\seveni  \scriptscriptfont1=\fivei%
  \textfont2=\tensy \scriptfont2=\sevensy \scriptscriptfont2=\fivesy%
  \textfont\itfam=\tenit\def\it{\fam\itfam\tenit}%
  \ifprod@font
    \scriptfont\itfam=\sevenit
      \scriptscriptfont\itfam=\fiveit
  \else
    \scriptfont\itfam=\tenit
      \scriptscriptfont\itfam=\tenit
  \fi
  \textfont\bffam=\tenbf%
    \scriptfont\bffam=\sevenbf%
      \scriptscriptfont\bffam=\fivebf%
  \def\bf{\fam\bffam\tenbf}%
  \textfont\slfam=\tensl\def\sl{\fam\slfam\tensl}%
  \ifprod@font
    \scriptfont\slfam=\sevensl
      \scriptscriptfont\slfam=\fivesl
  \else
    \scriptfont\slfam=\tensl
      \scriptscriptfont\slfam=\tensl
  \fi
  \textfont\ttfam=\tentt\def\tt{\fam\ttfam\tentt}%
  \ifprod@font
    \scriptfont\ttfam=\seventt
      \scriptscriptfont\ttfam=\fivett
  \else
    \scriptfont\ttfam=\tentt
      \scriptscriptfont\ttfam=\tentt
  \fi
  \textfont\scfam=\tencsc\def\sc{\fam\scfam\tencsc}%
  \ifprod@font
    \scriptfont\scfam=\sevencsc
      \scriptscriptfont\scfam=\fivecsc
  \else
    \scriptfont\scfam=\tencsc
      \scriptscriptfont\scfam=\tencsc
  \fi
  \textfont\sffam=\tensf\def\sf{\fam\sffam\tensf}%
  \ifprod@font
    \scriptfont\sffam=\sevensf
      \scriptscriptfont\sffam=\fivesf
  \else
    \scriptfont\sffam=\tensf
      \scriptscriptfont\sffam=\tensf
  \fi
  \def\oldstyle{\fam\@ne\teni}%
  \b@ls{11pt}\rm\@xpt%
}
\def\@xpt{}

\def\elevenpoint{
  \def\rm{\fam0\elevenrm}%
  \textfont0=\elevenrm \scriptfont0=\eightrm \scriptscriptfont0=\sixrm%
  \textfont1=\eleveni  \scriptfont1=\eighti  \scriptscriptfont1=\sixi%
  \textfont2=\elevensy \scriptfont2=\eightsy \scriptscriptfont2=\sixsy%
  \textfont\itfam=\elevenit\def\it{\fam\itfam\elevenit}%
  \ifprod@font
    \scriptfont\itfam=\eightit
      \scriptscriptfont\itfam=\sixit
  \else
    \scriptfont\itfam=\elevenit
      \scriptscriptfont\itfam=\elevenit
  \fi
  \textfont\bffam=\elevenbf%
    \scriptfont\bffam=\eightbf%
      \scriptscriptfont\bffam=\sixbf%
  \def\bf{\fam\bffam\elevenbf}%
  \textfont\slfam=\elevensl\def\sl{\fam\slfam\elevensl}%
  \ifprod@font
    \scriptfont\slfam=\eightsl
      \scriptscriptfont\slfam=\sixsl
  \else
    \scriptfont\slfam=\elevensl
      \scriptscriptfont\slfam=\elevensl
  \fi
  \textfont\ttfam=\eleventt\def\tt{\fam\ttfam\eleventt}%
  \ifprod@font
    \scriptfont\ttfam=\eighttt
      \scriptscriptfont\ttfam=\sixtt
  \else
    \scriptfont\ttfam=\eleventt
      \scriptscriptfont\ttfam=\eleventt
  \fi
  \textfont\scfam=\elevencsc\def\sc{\fam\scfam\elevencsc}%
  \ifprod@font
    \scriptfont\scfam=\eightcsc
      \scriptscriptfont\scfam=\sixcsc
  \else
    \scriptfont\scfam=\elevencsc
      \scriptscriptfont\scfam=\elevencsc
  \fi
  \textfont\sffam=\elevensf\def\sf{\fam\sffam\elevensf}%
  \ifprod@font
    \scriptfont\sffam=\eightsf
      \scriptscriptfont\sffam=\sixsf
  \else
    \scriptfont\sffam=\elevensf
      \scriptscriptfont\sffam=\elevensf
  \fi
  \def\oldstyle{\fam\@ne\eleveni}%
  \b@ls{13pt}\rm\@xipt%
}
\def\@xipt{}

\def\fourteenpoint{
  \def\rm{\fam0\fourteenrm}%
  \textfont0\fourteenrm  \scriptfont0\tenrm  \scriptscriptfont0\sevenrm%
  \textfont1\fourteeni   \scriptfont1\teni   \scriptscriptfont1\seveni%
  \textfont2\fourteensy  \scriptfont2\tensy  \scriptscriptfont2\sevensy%
  \textfont\itfam=\fourteenit\def\it{\fam\itfam\fourteenit}%
  \ifprod@font
    \scriptfont\itfam=\tenit
      \scriptscriptfont\itfam=\sevenit
  \else
    \scriptfont\itfam=\fourteenit
      \scriptscriptfont\itfam=\fourteenit
  \fi
  \textfont\bffam=\fourteenbf%
    \scriptfont\bffam=\tenbf%
      \scriptscriptfont\bffam=\sevenbf%
  \def\bf{\fam\bffam\fourteenbf}%
  \textfont\slfam=\fourteensl\def\sl{\fam\slfam\fourteensl}%
  \ifprod@font
    \scriptfont\slfam=\tensl
      \scriptscriptfont\slfam=\sevensl
  \else
    \scriptfont\slfam=\fourteensl
      \scriptscriptfont\slfam=\fourteensl
  \fi
  \textfont\ttfam=\fourteentt\def\tt{\fam\ttfam\fourteentt}%
  \ifprod@font
    \scriptfont\ttfam=\tentt
      \scriptscriptfont\ttfam=\seventt
  \else
    \scriptfont\ttfam=\fourteentt
      \scriptscriptfont\ttfam=\fourteentt
  \fi
  \textfont\scfam=\fourteencsc\def\sc{\fam\scfam\fourteencsc}%
  \ifprod@font
    \scriptfont\scfam=\tencsc
      \scriptscriptfont\scfam=\sevencsc
  \else
    \scriptfont\scfam=\fourteencsc
      \scriptscriptfont\scfam=\fourteencsc
  \fi
  \textfont\sffam=\fourteensf\def\sf{\fam\sffam\fourteensf}%
  \ifprod@font
    \scriptfont\sffam=\tensf
      \scriptscriptfont\sffam=\sevensf
  \else
    \scriptfont\sffam=\fourteensf
      \scriptscriptfont\sffam=\fourteensf
  \fi
  \def\oldstyle{\fam\@ne\fourteeni}%
  \b@ls{17pt}\rm\@xivpt%
}
\def\@xivpt{}

\def\seventeenpoint{
  \def\rm{\fam0\seventeenrm}%
  \textfont0\seventeenrm  \scriptfont0\twelverm  \scriptscriptfont0\tenrm%
  \textfont1\seventeeni   \scriptfont1\twelvei   \scriptscriptfont1\teni%
  \textfont2\seventeensy  \scriptfont2\twelvesy  \scriptscriptfont2\tensy%
  \textfont\itfam=\seventeenit\def\it{\fam\itfam\seventeenit}%
  \ifprod@font
    \scriptfont\itfam=\twelveit
      \scriptscriptfont\itfam=\tenit
  \else
    \scriptfont\itfam=\seventeenit
      \scriptscriptfont\itfam=\seventeenit
  \fi
  \textfont\bffam=\seventeenbf%
    \scriptfont\bffam=\twelvebf%
      \scriptscriptfont\bffam=\tenbf%
  \def\bf{\fam\bffam\seventeenbf}%
  \textfont\slfam=\seventeensl\def\sl{\fam\slfam\seventeensl}%
  \ifprod@font
    \scriptfont\slfam=\twelvesl
      \scriptscriptfont\slfam=\tensl
  \else
    \scriptfont\slfam=\seventeensl
      \scriptscriptfont\slfam=\seventeensl
  \fi
  \textfont\ttfam=\seventeentt\def\tt{\fam\ttfam\seventeentt}%
  \ifprod@font
    \scriptfont\ttfam=\twelvett
      \scriptscriptfont\ttfam=\tentt
  \else
    \scriptfont\ttfam=\seventeentt
      \scriptscriptfont\ttfam=\seventeentt
  \fi
  \textfont\scfam=\seventeencsc\def\sc{\fam\scfam\seventeencsc}%
  \ifprod@font
    \scriptfont\scfam=\twelvecsc
      \scriptscriptfont\scfam=\tencsc
  \else
    \scriptfont\scfam=\seventeencsc
      \scriptscriptfont\scfam=\seventeencsc
  \fi
  \textfont\sffam=\seventeensf\def\sf{\fam\sffam\seventeensf}%
  \ifprod@font
    \scriptfont\sffam=\twelvesf
      \scriptscriptfont\sffam=\tensf
  \else
    \scriptfont\sffam=\seventeensf
      \scriptscriptfont\sffam=\seventeensf
  \fi
  \def\oldstyle{\fam\@ne\seventeeni}%
  \b@ls{20pt}\rm\@xviipt%
}
\def\@xviipt{}

\lineskip=1pt      \normallineskip=\lineskip
\lineskiplimit=\z@ \normallineskiplimit=\lineskiplimit


\def\,{\relax\ifmmode \mskip\thinmuskip\else \thinspace\fi}
\let\protect=\relax

\long\def\@ifundefined#1#2#3{\expandafter\ifx\csname
  #1\endcsname\relax#2\else#3\fi}




\newtoks\math@groups \math@groups={}
\def\addtom@thgroup#1#2{#1\expandafter{\the#1#2}} 



\def\addtosizeh@ok#1#2#3#4{%
  \expandafter\def\csname @#1pt\endcsname{%
    \def\s@ze{#2}\def\ss@ze{#3}\def\sss@ze{#4}\the\math@groups%
  }%
}



\let\resetsizehook=\addtosizeh@ok


\ifprod@font
  \addtosizeh@ok{viii} {8} {6}  {5}
  \addtosizeh@ok{ix}   {9} {6}  {5}
  \addtosizeh@ok{x}    {10}{7}  {5}
  \addtosizeh@ok{xi}   {11}{8}  {6}
  \addtosizeh@ok{xiv}  {14}{10} {7}
  \addtosizeh@ok{xvii} {17}{12}{10}
\else
  \addtosizeh@ok{viii} {8}     {6}     {5}
  \addtosizeh@ok{ix}   {9}     {6}     {5}
  \addtosizeh@ok{x}    {10}    {7}     {5}
  \addtosizeh@ok{xi}   {10.95} {8}     {6}
  \addtosizeh@ok{xiv}  {14.4}  {10}    {7}
  \addtosizeh@ok{xvii} {17.28} {12}    {10}
\fi

\def\get@font#1#2#3{%
  \edef\fonts@ze{\romannumeral#3}
  \edef\fontn@me{\fonts@ze#1}
  \@ifundefined{\fontn@me}%
    {
     \global\expandafter\font\csname \fontn@me\endcsname=#2 at #3pt}%
    {}%
}

\def\ass@tfont#1#2{%
  \xdef\fam@name{\csname #1\endcsname}%
  \xdef\font@name{\csname #2\endcsname}%
  \let\textfont@name\font@name
  \textfont\fam@name\textfont@name
}

\def\ass@sfont#1#2{%
  \xdef\fam@name{\csname #1\endcsname}%
  \xdef\font@name{\csname #2\endcsname}%
  \let\textfont@name\font@name
  \scriptfont\fam@name\textfont@name
}

\def\ass@ssfont#1#2{%
  \xdef\fam@name{\csname #1\endcsname}%
  \xdef\font@name{\csname #2\endcsname}%
  \let\textfont@name\font@name
  \scriptscriptfont\fam@name\textfont@name
}


\def\NewSymbolFont#1#2{%
  \expandafter\ifx\csname sym#1fam\endcsname\relax 
    \expandafter\newfam\csname sym#1fam\endcsname
    \expandafter\edef\csname sym#1fam\endcsname{\the\allocationnumber}%
    \addtom@thgroup\math@groups{%
      \get@font{#1}{#2}{\s@ze}%
      \ass@tfont{sym#1fam}{\fontn@me}%
      \get@font{#1}{#2}{\ss@ze}%
      \ass@sfont{sym#1fam}{\fontn@me}%
      \get@font{#1}{#2}{\sss@ze}%
      \ass@ssfont{sym#1fam}{\fontn@me}%
    }%
  \else
    \errmessage{Family `#1' already defined}%
  \fi
}


\def\NewMathSymbol#1#2#3#4{%
  \edef\f@mly{\expandafter\hexnumber{\csname sym#3fam\endcsname}}%
  \mathchardef#1="#2\f@mly#4\relax
}


\newif\ifd@f

\def\NewMathDelimiter#1#2#3#4#5#6{%
  \d@ftrue
  \expandafter\ifx\csname sym#3fam\endcsname\relax
    \d@ffalse \errmessage{Family `#3' is not defined}%
  \fi
  \expandafter\ifx\csname sym#5fam\endcsname\relax
    \d@ffalse \errmessage{Family `#5' is not defined}%
  \fi
  \ifd@f
    \edef\f@mly{\expandafter\hexnumber{\csname sym#3fam\endcsname}}%
    \edef\f@mlytw@{\expandafter\hexnumber{\csname sym#5fam\endcsname}}%
    \xdef#1{\delimiter"#2\f@mly #4\f@mlytw@ #6\relax}%
  \fi
}


\def\setboxz@h{\setbox\z@\hbox}
\def\wdz@{\wd\z@}
\def\boxz@{\box\z@}
\def\setbox@ne{\setbox\@ne}
\def\wd@ne{\wd\@ne}

\def\math@atom#1#2{%
   \binrel@{#1}\binrel@@{#2}}
\def\binrel@#1{\setboxz@h{\thinmuskip0mu
  \medmuskip\m@ne mu\thickmuskip\@ne mu$#1\m@th$}%
 \setbox@ne\hbox{\thinmuskip0mu\medmuskip\m@ne mu\thickmuskip
  \@ne mu${}#1{}\m@th$}%
 \setbox\tw@\hbox{\hskip\wd@ne\hskip-\wdz@}}
\def\binrel@@#1{\ifdim\wd2<\z@\mathbin{#1}\else\ifdim\wd\tw@>\z@
 \mathrel{#1}\else{#1}\fi\fi}

\def\m@thit{1}

\def\set@skchar#1{\global\expandafter\skewchar
  \csname\fontn@me\endcsname=#1\relax}

\def\NewMathAlphabet#1#2#3{%
  \def\tst{#3}%
  \ifx\tst\empty\else 
    \expandafter\gdef\csname #1@sc\endcsname{}
  \fi
  \expandafter\def\csname #1\endcsname{
    \protect\csname @#1\endcsname}%
  \expandafter\def\csname @#1\endcsname##1{
    {%
    \begingroup
      \get@font{#1}{#2}{\s@ze}%
      \@ifundefined{#1@sc}{}{\set@skchar{#3}}%
      \ass@tfont{m@thit}{\fontn@me}%
      \get@font{#1}{#2}{\ss@ze}%
      \@ifundefined{#1@sc}{}{\set@skchar{#3}}%
      \ass@sfont{m@thit}{\fontn@me}%
      \get@font{#1}{#2}{\sss@ze}%
      \@ifundefined{#1@sc}{}{\set@skchar{#3}}%
      \ass@ssfont{m@thit}{\fontn@me}%
      \math@atom{##1}{%
      \mathchoice%
        {\hbox{$\m@th\displaystyle##1$}}%
        {\hbox{$\m@th\textstyle##1$}}%
        {\hbox{$\m@th\scriptstyle##1$}}%
        {\hbox{$\m@th\scriptscriptstyle##1$}}}%
    \endgroup
    }%
  }%
}


\newif\iffirstta  \firsttatrue

\def\set@hchar#1{\global\expandafter\hyphenchar
  \csname\fontn@me\endcsname=#1\relax}

\def\NewTextAlphabet#1#2#3{%
  \iffirstta
    \global\firsttafalse
    \newfam\scratchfam
    \edef\scrt@fam{\the\allocationnumber}%
  \fi
  \def\tst{#3}%
  \ifx\tst\empty\else 
    \expandafter\gdef\csname #1@hc\endcsname{}
  \fi
  \expandafter\def\csname #1\endcsname{
    \protect\csname t@#1\endcsname}%
  \long\expandafter\def\csname t@#1\endcsname##1{
    \ifmmode
      \typeout{Warning: do not use \expandafter\string\csname #1\endcsname
        \space in math mode}\fi%
    {%
      \get@font{#1}{#2}{\s@ze}\let\t@xtfnt=\fontn@me\relax
      \@ifundefined{#1@hc}{}{\set@hchar{#3}}%
      \ass@tfont{scrt@fam}{\fontn@me}%
      \get@font{#1}{#2}{\ss@ze}%
      \@ifundefined{#1@hc}{}{\set@hchar{#3}}%
      \ass@sfont{scrt@fam}{\fontn@me}%
      \get@font{#1}{#2}{\sss@ze}%
      \@ifundefined{#1@hc}{}{\set@hchar{#3}}%
      \ass@ssfont{scrt@fam}{\fontn@me}%
      \fam\scratchfam\csname\t@xtfnt\endcsname
    ##1%
    }%
  }%
  \expandafter\def\csname #1shape
    \endcsname{\protect\csname @#1shape\endcsname}%
  \expandafter\def\csname @#1shape\endcsname{
    \ifmmode
      \typeout{Warning: do not use \expandafter\string\csname
        #1shape\endcsname \space in math mode}\fi
      \get@font{#1}{#2}{\s@ze}\let\t@xtfnt=\fontn@me\relax
      \@ifundefined{#1@hc}{}{\set@hchar{#3}}%
      \ass@tfont{scrt@fam}{\fontn@me}%
      \get@font{#1}{#2}{\ss@ze}%
      \@ifundefined{#1@hc}{}{\set@hchar{#3}}%
      \ass@sfont{scrt@fam}{\fontn@me}%
      \get@font{#1}{#2}{\sss@ze}%
      \@ifundefined{#1@hc}{}{\set@hchar{#3}}%
      \ass@ssfont{scrt@fam}{\fontn@me}%
      \fam\scratchfam\csname\t@xtfnt\endcsname
  }%
}


\ifprod@font
  \def\math@itfnt{mtmib10}
  \def\math@syfnt{mtbsy10}
\else
  \def\math@itfnt{cmmib10}
  \def\math@syfnt{cmbsy10}
\fi

\def\m@thsy{2}

\def\bmath{\protect\@bmath}
\def\@bmath#1{%
  {%
  \begingroup
    \get@font{mthit}{\math@itfnt}{\s@ze}\set@skchar{'177}%
    \ass@tfont{m@thit}{\fontn@me}%
    \get@font{mthit}{\math@itfnt}{\ss@ze}\set@skchar{'177}%
    \ass@sfont{m@thit}{\fontn@me}%
    \get@font{mthit}{\math@itfnt}{\sss@ze}\set@skchar{'177}%
    \ass@ssfont{m@thit}{\fontn@me}%
    \get@font{mthsy}{\math@syfnt}{\s@ze}\set@skchar{'60}%
    \ass@tfont{m@thsy}{\fontn@me}%
    \get@font{mthsy}{\math@syfnt}{\ss@ze}\set@skchar{'60}%
    \ass@sfont{m@thsy}{\fontn@me}%
    \get@font{mthsy}{\math@syfnt}{\sss@ze}\set@skchar{'60}%
    \ass@ssfont{m@thsy}{\fontn@me}%
    \math@atom{#1}{%
    \mathchoice%
      {\hbox{$\m@th\displaystyle#1$}}%
      {\hbox{$\m@th\textstyle#1$}}%
      {\hbox{$\m@th\scriptstyle#1$}}%
      {\hbox{$\m@th\scriptscriptstyle#1$}}}%
  \endgroup
  }%
}



\def\ga{\mathrel{\mathchoice {\vcenter{\offinterlineskip\halign{\hfil
$\displaystyle##$\hfil\cr>\cr\sim\cr}}}
{\vcenter{\offinterlineskip\halign{\hfil$\textstyle##$\hfil\cr
>\cr\sim\cr}}}
{\vcenter{\offinterlineskip\halign{\hfil$\scriptstyle##$\hfil\cr
>\cr\sim\cr}}}
{\vcenter{\offinterlineskip\halign{\hfil$\scriptscriptstyle##$\hfil\cr
>\cr\sim\cr}}}}}

\def\diameter{{\ifmmode\mathchoice
{\ooalign{\hfil\hbox{$\displaystyle/$}\hfil\crcr
{\hbox{$\displaystyle\mathchar"20D$}}}}
{\ooalign{\hfil\hbox{$\textstyle/$}\hfil\crcr
{\hbox{$\textstyle\mathchar"20D$}}}}
{\ooalign{\hfil\hbox{$\scriptstyle/$}\hfil\crcr
{\hbox{$\scriptstyle\mathchar"20D$}}}}
{\ooalign{\hfil\hbox{$\scriptscriptstyle/$}\hfil\crcr
{\hbox{$\scriptscriptstyle\mathchar"20D$}}}}
\else{\ooalign{\hfil/\hfil\crcr\mathhexbox20D}}%
\fi}}

\def\sq{\ifmmode\squareforqed\else{\unskip\nobreak\hfil
\penalty50\hskip1em\null\nobreak\hfil\squareforqed
\parfillskip=0pt\finalhyphendemerits=0\endgraf}\fi}
\def\squareforqed{\hbox{\rlap{$\sqcap$}$\sqcup$}}


\def\bbbc{{\mathchoice {\setbox0=\hbox{$\displaystyle\rm C$}\hbox{\hbox
to0pt{\kern0.4\wd0\vrule height0.9\ht0\hss}\box0}}
{\setbox0=\hbox{$\textstyle\rm C$}\hbox{\hbox
to0pt{\kern0.4\wd0\vrule height0.9\ht0\hss}\box0}}
{\setbox0=\hbox{$\scriptstyle\rm C$}\hbox{\hbox
to0pt{\kern0.4\wd0\vrule height0.9\ht0\hss}\box0}}
{\setbox0=\hbox{$\scriptscriptstyle\rm C$}\hbox{\hbox
to0pt{\kern0.4\wd0\vrule height0.9\ht0\hss}\box0}}}}
\def\bbbq{{\mathchoice {\setbox0=\hbox{$\displaystyle\rm
Q$}\hbox{\raise
0.15\ht0\hbox to0pt{\kern0.4\wd0\vrule height0.8\ht0\hss}\box0}}
{\setbox0=\hbox{$\textstyle\rm Q$}\hbox{\raise
0.15\ht0\hbox to0pt{\kern0.4\wd0\vrule height0.8\ht0\hss}\box0}}
{\setbox0=\hbox{$\scriptstyle\rm Q$}\hbox{\raise
0.15\ht0\hbox to0pt{\kern0.4\wd0\vrule height0.7\ht0\hss}\box0}}
{\setbox0=\hbox{$\scriptscriptstyle\rm Q$}\hbox{\raise
0.15\ht0\hbox to0pt{\kern0.4\wd0\vrule height0.7\ht0\hss}\box0}}}}
\def\bbbt{{\mathchoice {\setbox0=\hbox{$\displaystyle\rm
T$}\hbox{\hbox to0pt{\kern0.3\wd0\vrule height0.9\ht0\hss}\box0}}
{\setbox0=\hbox{$\textstyle\rm T$}\hbox{\hbox
to0pt{\kern0.3\wd0\vrule height0.9\ht0\hss}\box0}}
{\setbox0=\hbox{$\scriptstyle\rm T$}\hbox{\hbox
to0pt{\kern0.3\wd0\vrule height0.9\ht0\hss}\box0}}
{\setbox0=\hbox{$\scriptscriptstyle\rm T$}\hbox{\hbox
to0pt{\kern0.3\wd0\vrule height0.9\ht0\hss}\box0}}}}
\def\bbbs{{\mathchoice
{\setbox0=\hbox{$\displaystyle     \rm S$}\hbox{\raise0.5\ht0\hbox
to0pt{\kern0.35\wd0\vrule height0.45\ht0\hss}\hbox
to0pt{\kern0.55\wd0\vrule height0.5\ht0\hss}\box0}}
{\setbox0=\hbox{$\textstyle        \rm S$}\hbox{\raise0.5\ht0\hbox
to0pt{\kern0.35\wd0\vrule height0.45\ht0\hss}\hbox
to0pt{\kern0.55\wd0\vrule height0.5\ht0\hss}\box0}}
{\setbox0=\hbox{$\scriptstyle      \rm S$}\hbox{\raise0.5\ht0\hbox
to0pt{\kern0.35\wd0\vrule height0.45\ht0\hss}\raise0.05\ht0\hbox
to0pt{\kern0.5\wd0\vrule height0.45\ht0\hss}\box0}}
{\setbox0=\hbox{$\scriptscriptstyle\rm S$}\hbox{\raise0.5\ht0\hbox
to0pt{\kern0.4\wd0\vrule height0.45\ht0\hss}\raise0.05\ht0\hbox
to0pt{\kern0.55\wd0\vrule height0.45\ht0\hss}\box0}}}}
\def\bbbz{{\mathchoice {\hbox{$\sf\textstyle Z\kern-0.4em Z$}}
{\hbox{$\sf\textstyle Z\kern-0.4em Z$}}
{\hbox{$\sf\scriptstyle Z\kern-0.3em Z$}}
{\hbox{$\sf\scriptscriptstyle Z\kern-0.2em Z$}}}}


\def\Nulle{0} 
\def\Afe{1}   
\def\Hae{2}   
\def\Hbe{3}   
\def\Hce{4}   
\def\Hde{5}   


\newcount\LastMac       \LastMac=\Nulle

\newskip\half      \half=5.5pt plus 1.5pt minus 2.25pt
\newskip\one       \one=11pt plus 3pt minus 5.5pt
\newskip\onehalf   \onehalf=16.5pt plus 5.5pt minus 8.25pt
\newskip\two       \two=22pt plus 5.5pt minus 11pt

\def\Half{\addvspace{\half}}
\def\One{\addvspace{\one}}
\def\OneHalf{\addvspace{\onehalf}}
\def\Two{\addvspace{\two}}

\def\Raggedright{
  \rightskip=\z@ plus \hsize\relax
}

\def\Fullout{
  \rightskip=\z@\relax
}

\def\Hang#1#2{
  \hangindent=#1%
  \hangafter=#2\relax
}


\newif\ifsp@page
\def\pagestyle#1{\csname ps@#1\endcsname}
\def\thispagestyle#1{\global\sp@pagetrue\gdef\sp@type{#1}}

\def\ps@titlepage{%
  \def\@oddhead{\eightpoint\noindent \the\CatchLine
    \ifprod@font\else\qquad Printed\ \today\qquad
      (MN plain \TeX\ macros\ v\@version)\fi \hfil}%
  \let\@evenhead=\@oddhead
  \def\@oddfoot{\eightpoint\copyright\ \@pubyear\ RAS\hfil}%
  \def\@evenfoot{\hfil\eightpoint\noindent\copyright\ \@pubyear\ RAS}%
}

\def\ps@headings{%
  \def\@oddhead{\elevenpoint\it\noindent
    \hfill\the\RightHeader\hskip1.5em\rm\folio}%
  \def\@evenhead{\elevenpoint\noindent
    \folio\hskip1.5em\it\the\LeftHeader\hfill}%
  \def\@oddfoot{\eightpoint\noindent\copyright\ \@pubyear\ RAS,
    MNRAS {\bf \@volume}, \@pagerange\hfil}%
  \def\@evenfoot{\hfil\eightpoint\copyright\ \@pubyear\ RAS,
    MNRAS {\bf \@volume}, \@pagerange}%
}

\def\ps@plate{%
  \def\@oddhead{\eightpoint\noindent\plt@cap\hfil}%
  \def\@evenhead{\eightpoint\noindent\plt@cap\hfil}%
  \def\@oddfoot{\eightpoint\noindent\copyright\ \@pubyear\ RAS,
    MNRAS {\bf \@volume}, \@pagerange\hfil}%
  \def\@evenfoot{\hfil\eightpoint\copyright\ \@pubyear\ RAS,
    MNRAS {\bf \@volume}, \@pagerange}%
}



\def\title#1{
  \bgroup
    \vbox to 8pt{\vss}%
    \seventeenpoint
    \Raggedright
    \noindent \strut{\bf #1}\par
  \egroup
}

\def\author#1{
  \bgroup
    \ifnum\LastMac=\Afe \OneHalf\else \vskip 21pt\fi
    \fourteenpoint
    \Raggedright
    \noindent \strut #1\par
    \vskip 3pt%
  \egroup
}

\def\affiliation#1{
  \bgroup
    \vskip -4pt%
    \eightpoint
    \Raggedright
    \noindent \strut {\it #1}\par
  \egroup
  \LastMac=\Afe\relax
}

\def\acceptedline#1{
  \bgroup
    \Two
    \eightpoint
    \Raggedright
    \noindent \strut #1\par
  \egroup
}

\long\def\abstract#1{%
  \bgroup
    \vskip 20pt%
    \leftskip 11pc\rightskip\z@
    \noindent{\ninebf ABSTRACT}\par
    \tenpoint
    \Fullout
    \noindent #1\par
  \egroup
}

\long\def\keywords#1{
  \bgroup
    \Half
    \leftskip 11pc\rightskip\z@
    \tenpoint
    \Fullout
    \noindent\hbox{\bf Key words:}\ #1\par
  \egroup
}


\def\maketitle{%
  \EndOpening
  \ifsinglecol \else \MakePage\fi
}


\def\pageoffset#1#2{\hoffset=#1\relax\voffset=#2\relax}


\def\@nameuse#1{\csname #1\endcsname}
\def\arabic#1{\@arabic{\@nameuse{#1}}}
\def\alph#1{\@alph{\@nameuse{#1}}}
\def\Alph#1{\@Alph{\@nameuse{#1}}}
\def\@arabic#1{\number #1}
\def\@Alph#1{\ifcase#1\or A\or B\or C\or D\else\@Ialph{#1}\fi}
\def\@Ialph#1{\ifcase#1\or \or \or \or \or E\or F\or G\or H\or I\or J\or
   K\or L\or M\or N\or O\or P\or Q\or R\or S\or T\or U\or V\or W\or X\or
   Y\or Z\else\errmessage{Counter out of range}\fi}
\def\@alph#1{\ifcase#1\or a\or b\or c\or d\else\@ialph{#1}\fi}
\def\@ialph#1{\ifcase#1\or \or \or \or \or e\or f\or g\or h\or i\or j\or
   k\or l\or m\or n\or o\or p\or q\or r\or s\or t\or u\or v\or w\or x\or y\or
   z\else\errmessage{Counter out of range}\fi}


\newcount\Eqnno
\newcount\SubEqnno

\def\theeq{\arabic{Eqnno}}
\def\thesubeq{\alph{SubEqnno}}

\def\stepeq{\relax
  \global\SubEqnno \z@
  \global\advance\Eqnno \@ne\relax
  {\rm (\theeq)}%
}

\def\startsubeq{\relax
  \global\SubEqnno \z@
  \global\advance\Eqnno \@ne\relax
  \stepsubeq
}

\def\stepsubeq{\relax
  \global\advance\SubEqnno \@ne\relax
  {\rm (\theeq\thesubeq)}%
}


\newcount\Sec        
\newcount\SecSec
\newcount\SecSecSec

\def\thesection{\arabic{Sec}}
\def\thesubsection{\thesection.\arabic{SecSec}}
\def\thesubsubsection{\thesubsection.\arabic{SecSecSec}}

\Sec=\z@

\def\:{\let\@sptoken= } \:  
\def\:{\@xifnch} \expandafter\def\: {\futurelet\@tempc\@ifnch}

\def\@ifnextchar#1#2#3{%
  \let\@tempMACe #1%
  \def\@tempMACa{#2}%
  \def\@tempMACb{#3}%
  \futurelet \@tempMACc\@ifnch%
}

\def\@ifnch{%
\ifx \@tempMACc \@sptoken%
  \let\@tempMACd\@xifnch%
\else%
  \ifx \@tempMACc \@tempMACe%
    \let\@tempMACd\@tempMACa%
  \else%
    \let\@tempMACd\@tempMACb%
  \fi%
\fi%
\@tempMACd%
}

\def\@ifstar#1#2{\@ifnextchar *{\def\@tempMACa*{#1}\@tempMACa}{#2}}

\newskip\@tempskipb

\def\addvspace#1{%
  \ifvmode\else \endgraf\fi%
  \ifdim\lastskip=\z@%
    \vskip #1\relax%
  \else%
    \@tempskipb#1\relax\@xaddvskip%
  \fi%
}

\def\@xaddvskip{%
  \ifdim\lastskip<\@tempskipb%
    \vskip-\lastskip%
    \vskip\@tempskipb\relax%
  \else%
    \ifdim\@tempskipb<\z@%
      \ifdim\lastskip<\z@ \else%
        \advance\@tempskipb\lastskip%
        \vskip-\lastskip\vskip\@tempskipb%
      \fi%
    \fi%
  \fi%
}

\newskip\@tmpSKIP

\def\addpen#1{%
  \ifvmode
    \if@nobreak
    \else
      \ifdim\lastskip=\z@
        \penalty#1\relax
      \else
        \@tmpSKIP=\lastskip
        \vskip -\lastskip
        \penalty#1\vskip\@tmpSKIP
      \fi
    \fi
  \fi
}

\newcount\@clubpen   \@clubpen=\clubpenalty
\newif\if@nobreak    \@nobreakfalse

\def\@noafterindent{%
  \global\@nobreaktrue
  \everypar{\if@nobreak
              \global\@nobreakfalse
              \clubpenalty \@M
              {\setbox\z@\lastbox}%
              \LastMac=\Nulle\relax%
            \else
              \clubpenalty \@clubpen
              \everypar{}%
            \fi}%
}

\newcount\gds@cbrk   \gds@cbrk=-300

\def\@nohdbrk{\interlinepenalty \@M\relax}

\let\@par=\par
\def\@restorepar{\def\par{\@par}}

\newif\if@endpe   \@endpefalse
 
\def\@doendpe{\@endpetrue \@nobreakfalse \LastMac=\Nulle\relax%
     \def\par{\@restorepar\everypar{}\par\@endpefalse}%
              \everypar{\setbox\z@\lastbox\everypar{}\@endpefalse}%
}

\def\section{\@ifstar{\@ssection}{\@section}}

\def\@section#1{
  \if@nobreak
    \everypar{}%
    \ifnum\LastMac=\Hae \addvspace{\half}\fi
  \else
    \addpen{\gds@cbrk}%
    \addvspace{\two}%
  \fi
  \bgroup
    \ninepoint\bf
    \Raggedright
    \global\advance\Sec \@ne
    \ifappendix
      \global\Eqnno=\z@ \global\SubEqnno=\z@\relax
      \def\ch@ck{#1}%
      \ifx\ch@ck\empty \def\c@lon{}\else\def\c@lon{:}\fi
      \noindent\@nohdbrk APPENDIX\ \thesection\c@lon\hskip 0.5em%
        \uppercase{#1}\par
    \else
      \noindent\@nohdbrk\thesection\hskip 1pc \uppercase{#1}\par
    \fi
    \global\SecSec=\z@
  \egroup
  \nobreak
  \vskip\half
  \nobreak
  \@noafterindent
  \LastMac=\Hae\relax
}

\def\@ssection#1{
  \if@nobreak
    \everypar{}%
    \ifnum\LastMac=\Hae \addvspace{\half}\fi
  \else
    \addpen{\gds@cbrk}%
    \addvspace{\two}%
  \fi
  \bgroup
    \ninepoint\bf
    \Raggedright
    \noindent\@nohdbrk\uppercase{#1}\par
  \egroup
  \nobreak
  \vskip\half
  \nobreak
  \@noafterindent
  \LastMac=\Hae\relax
}

\def\subsection{\@ifstar{\@ssubsection}{\@subsection}}

\def\@subsection#1{
  \if@nobreak
    \everypar{}%
    \ifnum\LastMac=\Hae \addvspace{1pt plus 1pt minus .5pt}\fi
  \else
    \addpen{\gds@cbrk}%
    \addvspace{\onehalf}%
  \fi
  \bgroup
    \ninepoint\bf
    \Raggedright
    \global\advance\SecSec \@ne
    \noindent\@nohdbrk\thesubsection \hskip 1pc\relax #1\par
    \global\SecSecSec=\z@
  \egroup
  \nobreak
  \vskip\half
  \nobreak
  \@noafterindent
  \LastMac=\Hbe\relax
}

\def\@ssubsection#1{
  \if@nobreak
    \everypar{}%
    \ifnum\LastMac=\Hae \addvspace{1pt plus 1pt minus .5pt}\fi
  \else
    \addpen{\gds@cbrk}%
    \addvspace{\onehalf}%
  \fi
  \bgroup
    \ninepoint\bf
    \Raggedright
    \noindent\@nohdbrk #1\par
  \egroup
  \nobreak
  \vskip\half
  \nobreak
  \@noafterindent
  \LastMac=\Hbe\relax
}

\def\subsubsection{\@ifstar{\@ssubsubsection}{\@subsubsection}}

\def\@subsubsection#1{
  \if@nobreak
    \everypar{}%
    \ifnum\LastMac=\Hbe \addvspace{1pt plus 1pt minus .5pt}\fi
  \else
    \addpen{\gds@cbrk}%
    \addvspace{\onehalf}%
  \fi
  \bgroup
    \ninepoint\it
    \Raggedright
    \global\advance\SecSecSec \@ne
    \noindent\@nohdbrk\thesubsubsection \hskip 1pc\relax #1\par
  \egroup
  \nobreak
  \vskip\half
  \nobreak
  \@noafterindent
  \LastMac=\Hce\relax
}

\def\@ssubsubsection#1{
  \if@nobreak
    \everypar{}%
    \ifnum\LastMac=\Hbe \addvspace{1pt plus 1pt minus .5pt}\fi
  \else
    \addpen{\gds@cbrk}%
    \addvspace{\onehalf}%
  \fi
  \bgroup
    \ninepoint\it
    \Raggedright
    \noindent\@nohdbrk #1\par
  \egroup
  \nobreak
  \vskip\half
  \nobreak
  \@noafterindent
  \LastMac=\Hce\relax
}

\def\paragraph#1{
  \if@nobreak
    \everypar{}%
  \else
    \addpen{\gds@cbrk}%
    \addvspace{\one}%
  \fi%
  \bgroup%
    \ninepoint\it
    \noindent #1\ \nobreak%
  \egroup
  \LastMac=\Hde\relax
  \ignorespaces
}


\newif\ifappendix

\def\appendix{%
  \global\appendixtrue
  \def\thesection{\Alph{Sec}}%
  \def\thesubsection{\thesection\arabic{SecSec}}%
  \def\theeq{\thesection\arabic{Eqnno}}%
  \Sec=\z@ \SecSec=\z@ \SecSecSec=\z@ \Eqnno=\z@ \SubEqnno=\z@\relax
}




\def\beginlist{%
  \par\if@nobreak \else\addvspace{\half}\fi%
  \bgroup%
    \ninepoint
    \let\item=\list@item%
}

\def\list@item{%
  \par\noindent\hskip 1em\relax%
  \ignorespaces%
}

\def\endlist{\par\egroup\addvspace{\half}\@doendpe}


\def\beginrefs{%
  \par
  \bgroup
    \eightpoint
    \Fullout
    \let\bibitem=\bib@item
}

\def\bib@item{%
  \par\parindent=1.5em\Hang{1.5em}{1}%
  \everypar={\Hang{1.5em}{1}\ignorespaces}%
  \noindent\ignorespaces
}

\def\endrefs{\par\egroup\@doendpe}


\newtoks\CatchLine

\def\@journal{Mon.\ Not.\ R.\ Astron.\ Soc.\ }  
\def\@pubyear{1994}        
\def\@pagerange{000--000}  
\def\@volume{000}          
\def\@microfiche{}         %

\def\pubyear#1{\gdef\@pubyear{#1}\@makecatchline}
\def\pagerange#1{\gdef\@pagerange{#1}\@makecatchline}
\def\volume#1{\gdef\@volume{#1}\@makecatchline}
\def\microfiche#1{\gdef\@microfiche{and Microfiche\ #1}\@makecatchline}

\def\@makecatchline{%
  \global\CatchLine{%
    {\rm \@journal {\bf \@volume},\ \@pagerange\ (\@pubyear)\ \@microfiche}}%
}

\@makecatchline 

\newtoks\LeftHeader
\def\shortauthor#1{
  \global\LeftHeader{#1}%
}

\newtoks\RightHeader
\def\shorttitle#1{
  \global\RightHeader{#1}%
}

\def\PageHead{
  \begingroup
    \ifsp@page
      \csname ps@\sp@type\endcsname
    \fi
    \ifodd\pageno
      \let\the@head=\@oddhead
    \else
      \let\the@head=\@evenhead
    \fi
    \vbox to \z@{\vskip-22.5\p@%
      \hbox to \PageWidth{\vbox to8.5\p@{}%
        \the@head
      }%
    \vss}%
  \endgroup
  \nointerlineskip
}

\gdef\PageFoot{%
  \nointerlineskip%
  \begingroup
  \ifsp@page
    \csname ps@\sp@type\endcsname
    \global\sp@pagefalse
  \fi
  \vbox to 22pt{\vfil%
    \hbox to \PageWidth{%
      \eightpoint\strut\noindent
      \ifodd\pageno
        \@oddfoot
      \else
        \@evenfoot
      \fi
    }%
  }%
  \endgroup
}

\def\today{%
  \number\day\space
  \ifcase\month\or January\or February\or March\or April\or May\or June\or
    July\or August\or September\or October\or November\or December\fi
  \space\number\year%
}

\def\authorcomment#1{%
  \gdef\PageFoot{%
    \nointerlineskip%
    \vbox to 20pt{\vfil%
      \hbox to \PageWidth{\elevenpoint\noindent \hfil #1 \hfil}}%
  }%
}


\newif\ifplate@page
\newbox\plt@box

\def\beginplatepage{%
  \let\plate=\plate@head
  \let\caption=\fig@caption
  \global\setbox\plt@box=\vbox\bgroup
  \TEMPDIMEN=\PageWidth 
  \hsize=\PageWidth\relax
}

\def\endplatepage{\par\egroup\global\plate@pagetrue}
\def\plate@head#1{\gdef\plt@cap{#1}}


\def\letters{%
  \gdef\folio{\ifnum\pageno<\z@ L\romannumeral-\pageno
    \else L\number\pageno \fi}%
}


\newdimen\mathindent

\global\mathindent=\z@
\global\everydisplay{\global\@dspwd=\displaywidth\displaysetup}


\def\@displaylines#1{
  {}$\displ@y\hbox{\vbox{\halign{$\@lign\hfil\displaystyle##\hfil$\crcr
  #1\crcr}}}${}%
}

\def\@eqalign#1{\null\vcenter{\openup\jot\m@th
  \ialign{\strut\hfil$\displaystyle{##}$&$\displaystyle{{}##}$\hfil
      \crcr#1\crcr}}%
}

\def\@eqalignno#1{
  \global\advance\@dspwd by -\mathindent%
  {}$\displ@y\hbox{\vbox{\halign to\@dspwd%
  {\hfil$\@lign\displaystyle{##}$\tabskip\z@skip
  &$\@lign\displaystyle{{}##}$\hfil\tabskip\centering
  &\llap{$\@lign##$}\tabskip\z@skip\crcr
  #1\crcr}}}${}%
}


\global\let\displaylines=\@displaylines
\global\let\eqalign=\@eqalign
\global\let\eqalignno=\@eqalignno
\global\let\leqalignno=\@eqalignno

\newdimen\@dspwd   \@dspwd=\z@
\newif\if@eqno
\newif\if@leqno
\newtoks\@eqn
\newtoks\@eq

\def\displaysetup#1$${\displaytest#1\eqno\eqno\displaytest}

\def\displaytest#1\eqno#2\eqno#3\displaytest{%
 \if!#3!\ldisplaytest#1\leqno\leqno\ldisplaytest
 \else\@eqnotrue\@leqnofalse\@eqn={#2}\@eq={#1}\fi
 \generaldisplay$$}

\def\ldisplaytest#1\leqno#2\leqno#3\ldisplaytest{%
\@eq={#1}%
 \if!#3!\@eqnofalse\else\@eqnotrue\@leqnotrue
  \@eqn={#2}\fi}

\def\generaldisplay{%
  \if@eqno
    \if@leqno
      \hbox to \displaywidth{\noindent
        \rlap{$\displaystyle\the\@eqn$}%
        \hskip\mathindent$\displaystyle\the\@eq$\hfil}%
    \else
      \hbox to \displaywidth{\noindent
        \hskip\mathindent
        $\displaystyle\the\@eq$\hfil$\displaystyle\the\@eqn$}%
    \fi
  \else
    \hbox to \displaywidth{\noindent
      \hskip\mathindent$\displaystyle\the\@eq$\hfil}%
  \fi
}


\def\@notice{%
  \par\Two%
  \noindent{\b@ls{11pt}\ninerm This paper has been produced using the
    Royal Astronomical Society/Blackwell Science \TeX\ macros.\par}%
}

\outer\def
\section*{References}
\beginrefs
\bibitem
Ali S., Rossinot P., Piccirillo L., Gear W. K., Mauskopf P.,
Ade P., Haynes V., Timbie P., 2002, in
De Petris M., Gervasi M., eds,
AIP Conf Proc. Vol. 66, Experimental Cosmology at Millimetre Wavelengths,
AIP, New York, p. 126
\bibitem
Bridle A. H., Schwab F. R., 1999, in
Taylor G. B., Carilli C. L., Perley R. A., eds,
ASP Conf Ser. 180, Synthesis Imaging in Radio Astronomy {\sc II},  
ASP, San Francisco, p. 371
\bibitem
Clark B. G., 1999, in 
Taylor G. B., Carilli C. L., Perley R. A., eds,
ASP Conf Ser. 180, Synthesis Imaging in Radio Astronomy {\sc II},  
ASP, San Francisco, p. 1
\bibitem
Cornwell T. J., Holdaway M. A., Uson J. M., 1993, A\&A, 271, 697
\
\bibitem
Cotton W. D., 1999, in 
Taylor G. B., Carilli C. L., Perley R. A., eds,
ASP Conf Ser. 180, Synthesis Imaging in Radio Astronomy {\sc II},  
ASP, San Francisco, p. 357
\bibitem
Ekers R. D., Rots A. H., 1979, in 
van Schooneveid C., eds,
Astrop. \& Sp. Sc. Lib. 76,
Image Formation from Coherence Functions in Astronomy,
Reidel, Dordrecht, p. 61
\bibitem
Rao A. P., Velusamy T., 1984, in
Roberts J. A., eds,
Indirect Imaging, 
Cambridge Univ. Press, Cambridge, p. 193
\bibitem
Subrahmanyan R. 2002, in 
Chen L. -W., Ma C. -P., Ng K. -W., Pen U. -L., eds,
ASP Conf Ser. 257, AMiBA 2001: High-z Clusters,
Missing Baryons, and CMB Polarization, ASP, San Francisco, p. 309
\bibitem
White M., Carlstrom J. E., Dragovan M., Holzapfel W., 1999, ApJ, 
514, 1
\endrefs
\bye{\@notice\par\vfill\supereject\end}


\def\start@mess{%
  Monthly notices of the RAS journal style (\@typeface)\space
    v\@version,\space \@verdate.%
}

\everyjob{\Warn{\start@mess}}



\newif\if@debug \@debugfalse  

\def\Print#1{\if@debug\immediate\write16{#1}\else \fi}
\def\Warn#1{\immediate\write16{#1}}
\def\wlog#1{}

\newcount\Iteration 

\def\Single{0} \def\Double{1}                 
\def\Figure{0} \def\Table{1}                  

\def\InStack{0}  
\def\InZoneA{1}
\def\InZoneB{2}
\def\InZoneC{3}

\newcount\TEMPCOUNT 
\newdimen\TEMPDIMEN 
\newbox\TEMPBOX     
\newbox\VOIDBOX     

\newcount\LengthOfStack 
\newcount\MaxItems      
\newcount\StackPointer
\newcount\Point         
\newcount\NextFigure    
\newcount\NextTable     
\newcount\NextItem      

\newcount\StatusStack   
\newcount\NumStack      
\newcount\TypeStack     
\newcount\SpanStack     
\newcount\BoxStack      

\newcount\ItemSTATUS    
\newcount\ItemNUMBER    
\newcount\ItemTYPE      
\newcount\ItemSPAN      
\newbox\ItemBOX         
\newdimen\ItemSIZE      

\newdimen\PageHeight    
\newdimen\TextLeading   
\newdimen\Feathering    
\newcount\LinesPerPage  
\newdimen\ColumnWidth   
\newdimen\ColumnGap     
\newdimen\PageWidth     
\newdimen\BodgeHeight   
\newcount\Leading       

\newdimen\ZoneBSize  
\newdimen\TextSize   
\newbox\ZoneABOX     
\newbox\ZoneBBOX     
\newbox\ZoneCBOX     

\newif\ifFirstSingleItem
\newif\ifFirstZoneA
\newif\ifMakePageInComplete
\newif\ifMoreFigures \MoreFiguresfalse 
\newif\ifMoreTables  \MoreTablesfalse  

\newif\ifFigInZoneB 
\newif\ifFigInZoneC 
\newif\ifTabInZoneB 
\newif\ifTabInZoneC

\newif\ifZoneAFullPage

\newbox\MidBOX    
\newbox\LeftBOX
\newbox\RightBOX
\newbox\PageBOX   

\newif\ifLeftCOL  
\LeftCOLtrue

\newdimen\ZoneBAdjust

\newcount\ItemFits
\def\Yes{1}
\def\No{2}


\MaxItems=15
\NextFigure=\z@        
\NextTable=\@ne

\BodgeHeight=6pt
\TextLeading=11pt    
\Leading=11
\Feathering=\z@      
\LinesPerPage=61     
\topskip=\TextLeading
\ColumnWidth=20pc    
\ColumnGap=2pc       

\newskip\ItemSepamount  
\ItemSepamount=\TextLeading plus \TextLeading minus 4pt

\parskip=\z@ plus .1pt
\parindent=18pt
\widowpenalty=\z@
\clubpenalty=10000
\tolerance=1500
\hbadness=1500
\abovedisplayskip=6pt plus 2pt minus 1pt
\belowdisplayskip=6pt plus 2pt minus 1pt
\abovedisplayshortskip=6pt plus 2pt minus 1pt
\belowdisplayshortskip=6pt plus 2pt minus 1pt

\frenchspacing

\ninepoint 

\PageHeight=682pt
\PageWidth=2\ColumnWidth
\advance\PageWidth by \ColumnGap

\pagestyle{headings}




\newcount\DUMMY \StatusStack=\allocationnumber
\newcount\DUMMY \newcount\DUMMY \newcount\DUMMY 
\newcount\DUMMY \newcount\DUMMY \newcount\DUMMY 
\newcount\DUMMY \newcount\DUMMY \newcount\DUMMY
\newcount\DUMMY \newcount\DUMMY \newcount\DUMMY 
\newcount\DUMMY \newcount\DUMMY \newcount\DUMMY

\newcount\DUMMY \NumStack=\allocationnumber
\newcount\DUMMY \newcount\DUMMY \newcount\DUMMY 
\newcount\DUMMY \newcount\DUMMY \newcount\DUMMY 
\newcount\DUMMY \newcount\DUMMY \newcount\DUMMY 
\newcount\DUMMY \newcount\DUMMY \newcount\DUMMY 
\newcount\DUMMY \newcount\DUMMY \newcount\DUMMY

\newcount\DUMMY \TypeStack=\allocationnumber
\newcount\DUMMY \newcount\DUMMY \newcount\DUMMY 
\newcount\DUMMY \newcount\DUMMY \newcount\DUMMY 
\newcount\DUMMY \newcount\DUMMY \newcount\DUMMY 
\newcount\DUMMY \newcount\DUMMY \newcount\DUMMY 
\newcount\DUMMY \newcount\DUMMY \newcount\DUMMY

\newcount\DUMMY \SpanStack=\allocationnumber
\newcount\DUMMY \newcount\DUMMY \newcount\DUMMY 
\newcount\DUMMY \newcount\DUMMY \newcount\DUMMY 
\newcount\DUMMY \newcount\DUMMY \newcount\DUMMY 
\newcount\DUMMY \newcount\DUMMY \newcount\DUMMY 
\newcount\DUMMY \newcount\DUMMY \newcount\DUMMY

\newbox\DUMMY   \BoxStack=\allocationnumber
\newbox\DUMMY   \newbox\DUMMY \newbox\DUMMY 
\newbox\DUMMY   \newbox\DUMMY \newbox\DUMMY 
\newbox\DUMMY   \newbox\DUMMY \newbox\DUMMY 
\newbox\DUMMY   \newbox\DUMMY \newbox\DUMMY 
\newbox\DUMMY   \newbox\DUMMY \newbox\DUMMY

\def\wlog{\immediate\write\m@ne}


\def\GetItemAll#1{%
 \GetItemSTATUS{#1}
 \GetItemNUMBER{#1}
 \GetItemTYPE{#1}
 \GetItemSPAN{#1}
 \GetItemBOX{#1}
}

\def\GetItemSTATUS#1{%
 \Point=\StatusStack
 \advance\Point by #1
 \global\ItemSTATUS=\count\Point
}

\def\GetItemNUMBER#1{%
 \Point=\NumStack
 \advance\Point by #1
 \global\ItemNUMBER=\count\Point
}

\def\GetItemTYPE#1{%
 \Point=\TypeStack
 \advance\Point by #1
 \global\ItemTYPE=\count\Point
}

\def\GetItemSPAN#1{%
 \Point\SpanStack
 \advance\Point by #1
 \global\ItemSPAN=\count\Point
}

\def\GetItemBOX#1{%
 \Point=\BoxStack
 \advance\Point by #1
 \global\setbox\ItemBOX=\vbox{\copy\Point}
 \global\ItemSIZE=\ht\ItemBOX
 \global\advance\ItemSIZE by \dp\ItemBOX
 \TEMPCOUNT=\ItemSIZE
 \divide\TEMPCOUNT by \Leading
 \divide\TEMPCOUNT by 65536
 \advance\TEMPCOUNT \@ne
 \ItemSIZE=\TEMPCOUNT pt
 \global\multiply\ItemSIZE by \Leading
}


\def\JoinStack{%
 \ifnum\LengthOfStack=\MaxItems 
  \Warn{WARNING: Stack is full...some items will be lost!}
 \else
  \Point=\StatusStack
  \advance\Point by \LengthOfStack
  \global\count\Point=\ItemSTATUS
  \Point=\NumStack
  \advance\Point by \LengthOfStack
  \global\count\Point=\ItemNUMBER
  \Point=\TypeStack
  \advance\Point by \LengthOfStack
  \global\count\Point=\ItemTYPE
  \Point\SpanStack
  \advance\Point by \LengthOfStack
  \global\count\Point=\ItemSPAN
  \Point=\BoxStack
  \advance\Point by \LengthOfStack
  \global\setbox\Point=\vbox{\copy\ItemBOX}
  \global\advance\LengthOfStack \@ne
  \ifnum\ItemTYPE=\Figure 
   \global\MoreFigurestrue
  \else
   \global\MoreTablestrue
  \fi
 \fi
}


\def\LeaveStack#1{%
 {\Iteration=#1
 \loop
 \ifnum\Iteration<\LengthOfStack
  \advance\Iteration \@ne
  \GetItemSTATUS{\Iteration}
   \advance\Point by \m@ne
   \global\count\Point=\ItemSTATUS
  \GetItemNUMBER{\Iteration}
   \advance\Point by \m@ne
   \global\count\Point=\ItemNUMBER
  \GetItemTYPE{\Iteration}
   \advance\Point by \m@ne
   \global\count\Point=\ItemTYPE
  \GetItemSPAN{\Iteration}
   \advance\Point by \m@ne
   \global\count\Point=\ItemSPAN
  \GetItemBOX{\Iteration}
   \advance\Point by \m@ne
   \global\setbox\Point=\vbox{\copy\ItemBOX}
 \repeat}
 \global\advance\LengthOfStack by \m@ne
}


\newif\ifStackNotClean

\def\CleanStack{%
 \StackNotCleantrue
 {\Iteration=\z@
  \loop
   \ifStackNotClean
    \GetItemSTATUS{\Iteration}
    \ifnum\ItemSTATUS=\InStack
     \advance\Iteration \@ne
     \else
      \LeaveStack{\Iteration}
    \fi
   \ifnum\LengthOfStack<\Iteration
    \StackNotCleanfalse
   \fi
 \repeat}
}


\def\FindItem#1#2{%
 \global\StackPointer=\m@ne 
 {\Iteration=\z@
  \loop
  \ifnum\Iteration<\LengthOfStack
   \GetItemSTATUS{\Iteration}
   \ifnum\ItemSTATUS=\InStack
    \GetItemTYPE{\Iteration}
    \ifnum\ItemTYPE=#1
     \GetItemNUMBER{\Iteration}
     \ifnum\ItemNUMBER=#2
      \global\StackPointer=\Iteration
      \Iteration=\LengthOfStack 
     \fi
    \fi
   \fi
  \advance\Iteration \@ne
 \repeat}
}


\def\FindNext{%
 \global\StackPointer=\m@ne 
 {\Iteration=\z@
  \loop
  \ifnum\Iteration<\LengthOfStack
   \GetItemSTATUS{\Iteration}
   \ifnum\ItemSTATUS=\InStack
    \GetItemTYPE{\Iteration}
   \ifnum\ItemTYPE=\Figure
    \ifMoreFigures
      \global\NextItem=\Figure
      \global\StackPointer=\Iteration
      \Iteration=\LengthOfStack 
    \fi
   \fi
   \ifnum\ItemTYPE=\Table
    \ifMoreTables
      \global\NextItem=\Table
      \global\StackPointer=\Iteration
      \Iteration=\LengthOfStack 
    \fi
   \fi
  \fi
  \advance\Iteration \@ne
 \repeat}
}


\def\ChangeStatus#1#2{%
 \Point=\StatusStack
 \advance\Point by #1
 \global\count\Point=#2
}



\def\Zone{\InZoneA}

\ZoneBAdjust=\z@

\def\MakePage{
 \global\ZoneBSize=\PageHeight
 \global\TextSize=\ZoneBSize
 \global\ZoneAFullPagefalse
 \global\topskip=\TextLeading
 \MakePageInCompletetrue
 \MoreFigurestrue
 \MoreTablestrue
 \FigInZoneBfalse
 \FigInZoneCfalse
 \TabInZoneBfalse
 \TabInZoneCfalse
 \global\FirstSingleItemtrue
 \global\FirstZoneAtrue
 \global\setbox\ZoneABOX=\box\VOIDBOX
 \global\setbox\ZoneBBOX=\box\VOIDBOX
 \global\setbox\ZoneCBOX=\box\VOIDBOX
 \loop
  \ifMakePageInComplete
 \FindNext
 \ifnum\StackPointer=\m@ne
  \NextItem=\m@ne
  \MoreFiguresfalse
  \MoreTablesfalse
 \fi
 \ifnum\NextItem=\Figure
   \FindItem{\Figure}{\NextFigure}
   \ifnum\StackPointer=\m@ne \global\MoreFiguresfalse
   \else
    \GetItemSPAN{\StackPointer}
    \ifnum\ItemSPAN=\Single \def\Zone{\InZoneB}\relax
     \ifFigInZoneC \global\MoreFiguresfalse\fi
    \else
     \def\Zone{\InZoneA}
     \ifFigInZoneB \def\Zone{\InZoneC}\fi
    \fi
   \fi
   \ifMoreFigures\Print{}\FigureItems\fi
 \fi
\ifnum\NextItem=\Table
   \FindItem{\Table}{\NextTable}
   \ifnum\StackPointer=\m@ne \global\MoreTablesfalse
   \else
    \GetItemSPAN{\StackPointer}
    \ifnum\ItemSPAN=\Single\relax
     \ifTabInZoneC \global\MoreTablesfalse\fi
    \else
     \def\Zone{\InZoneA}
     \ifTabInZoneB \def\Zone{\InZoneC}\fi
    \fi
   \fi
   \ifMoreTables\Print{}\TableItems\fi
 \fi
   \MakePageInCompletefalse 
   \ifMoreFigures\MakePageInCompletetrue\fi
   \ifMoreTables\MakePageInCompletetrue\fi
 \repeat
 \ifZoneAFullPage
  \global\TextSize=\z@
  \global\ZoneBSize=\z@
  \global\vsize=\z@\relax
  \global\topskip=\z@\relax
  \vbox to \z@{\vss}
  \eject
 \else
 \global\advance\ZoneBSize by -\ZoneBAdjust
 \global\vsize=\ZoneBSize
 \global\hsize=\ColumnWidth
 \global\ZoneBAdjust=\z@
 \ifdim\TextSize<23pt
 \Warn{}
 \Warn{* Making column fall short: TextSize=\the\TextSize *}
 \vskip-\lastskip\eject\fi
 \fi
}

\def\MakeRightCol{
 \global\TextSize=\ZoneBSize
 \MakePageInCompletetrue
 \MoreFigurestrue
 \MoreTablestrue
 \global\FirstSingleItemtrue
 \global\setbox\ZoneBBOX=\box\VOIDBOX
 \def\Zone{\InZoneB}
 \loop
  \ifMakePageInComplete
 \FindNext
 \ifnum\StackPointer=\m@ne
  \NextItem=\m@ne
  \MoreFiguresfalse
  \MoreTablesfalse
 \fi
 \ifnum\NextItem=\Figure
   \FindItem{\Figure}{\NextFigure}
   \ifnum\StackPointer=\m@ne \MoreFiguresfalse
   \else
    \GetItemSPAN{\StackPointer}
    \ifnum\ItemSPAN=\Double\relax
     \MoreFiguresfalse\fi
   \fi
   \ifMoreFigures\Print{}\FigureItems\fi
 \fi
 \ifnum\NextItem=\Table
   \FindItem{\Table}{\NextTable}
   \ifnum\StackPointer=\m@ne \MoreTablesfalse
   \else
    \GetItemSPAN{\StackPointer}
    \ifnum\ItemSPAN=\Double\relax
     \MoreTablesfalse\fi
   \fi
   \ifMoreTables\Print{}\TableItems\fi
 \fi
   \MakePageInCompletefalse 
   \ifMoreFigures\MakePageInCompletetrue\fi
   \ifMoreTables\MakePageInCompletetrue\fi
 \repeat
 \ifZoneAFullPage
  \global\TextSize=\z@
  \global\ZoneBSize=\z@
  \global\vsize=\z@\relax
  \global\topskip=\z@\relax
  \vbox to \z@{\vss}
  \eject
 \else
 \global\vsize=\ZoneBSize
 \global\hsize=\ColumnWidth
 \ifdim\TextSize<23pt
 \Warn{}
 \Warn{* Making column fall short: TextSize=\the\TextSize *}
 \vskip-\lastskip\eject\fi
\fi
}

\def\FigureItems{
 \Print{Considering...}
 \ShowItem{\StackPointer}
 \GetItemBOX{\StackPointer} 
 \GetItemSPAN{\StackPointer}
  \CheckFitInZone 
  \ifnum\ItemFits=\Yes
   \ifnum\ItemSPAN=\Single
     \ChangeStatus{\StackPointer}{\InZoneB} 
     \global\FigInZoneBtrue
     \ifFirstSingleItem
      \hbox{}\vskip-\BodgeHeight
     \global\advance\ItemSIZE by \TextLeading
     \fi
     \unvbox\ItemBOX\ItemSep
     \global\FirstSingleItemfalse
     \global\advance\TextSize by -\ItemSIZE
     \global\advance\TextSize by -\TextLeading
   \else
    \ifFirstZoneA
     \global\advance\ItemSIZE by \TextLeading
     \global\FirstZoneAfalse\fi
    \global\advance\TextSize by -\ItemSIZE
    \global\advance\TextSize by -\TextLeading
    \global\advance\ZoneBSize by -\ItemSIZE
    \global\advance\ZoneBSize by -\TextLeading
    \ifFigInZoneB\relax
     \else
     \ifdim\TextSize<3\TextLeading
     \global\ZoneAFullPagetrue
     \fi
    \fi
    \ChangeStatus{\StackPointer}{\Zone}
    \ifnum\Zone=\InZoneC \global\FigInZoneCtrue\fi
  \fi
   \Print{TextSize=\the\TextSize}
   \Print{ZoneBSize=\the\ZoneBSize}
  \global\advance\NextFigure \@ne
   \Print{This figure has been placed.}
  \else
   \Print{No space available for this figure...holding over.}
   \Print{}
   \global\MoreFiguresfalse
  \fi
}

\def\TableItems{
 \Print{Considering...}
 \ShowItem{\StackPointer}
 \GetItemBOX{\StackPointer} 
 \GetItemSPAN{\StackPointer}
  \CheckFitInZone 
  \ifnum\ItemFits=\Yes
   \ifnum\ItemSPAN=\Single
    \ChangeStatus{\StackPointer}{\InZoneB}
     \global\TabInZoneBtrue
     \ifFirstSingleItem
      \hbox{}\vskip-\BodgeHeight
     \global\advance\ItemSIZE by \TextLeading
     \fi
     \unvbox\ItemBOX\ItemSep
     \global\FirstSingleItemfalse
     \global\advance\TextSize by -\ItemSIZE
     \global\advance\TextSize by -\TextLeading
   \else
    \ifFirstZoneA
    \global\advance\ItemSIZE by \TextLeading
    \global\FirstZoneAfalse\fi
    \global\advance\TextSize by -\ItemSIZE
    \global\advance\TextSize by -\TextLeading
    \global\advance\ZoneBSize by -\ItemSIZE
    \global\advance\ZoneBSize by -\TextLeading
    \ifFigInZoneB\relax
     \else
     \ifdim\TextSize<3\TextLeading
     \global\ZoneAFullPagetrue
     \fi
    \fi
    \ChangeStatus{\StackPointer}{\Zone}
    \ifnum\Zone=\InZoneC \global\TabInZoneCtrue\fi
   \fi
  \global\advance\NextTable \@ne
   \Print{This table has been placed.}
  \else
  \Print{No space available for this table...holding over.}
   \Print{}
   \global\MoreTablesfalse
  \fi
}


\def\CheckFitInZone{%
{\advance\TextSize by -\ItemSIZE
 \advance\TextSize by -\TextLeading
 \ifFirstSingleItem
  \advance\TextSize by \TextLeading
 \fi
 \ifnum\Zone=\InZoneA\relax
  \else \advance\TextSize by -\ZoneBAdjust
 \fi
 \ifdim\TextSize<3\TextLeading \global\ItemFits=\No
 \else \global\ItemFits=\Yes\fi}
}

\def\BeginOpening{%
  \ninepoint
  \thispagestyle{titlepage}%
  \global\setbox\ItemBOX=\vbox\bgroup%
    \hsize=\PageWidth%
    \hrule height \z@
    \ifsinglecol\vskip 6pt\fi 
}

\let\begintopmatter=\BeginOpening  

\def\EndOpening{%
  \One
  \egroup
  \ifsinglecol
    \box\ItemBOX%
    \vskip\TextLeading plus 2\TextLeading
    \@noafterindent
  \else
    \ItemNUMBER=\z@%
    \ItemTYPE=\Figure
    \ItemSPAN=\Double
    \ItemSTATUS=\InStack
    \JoinStack
  \fi
}


\newif\if@here  \@herefalse

\def\no@float{\global\@heretrue}
\let\nofloat=\relax 

\def\beginfigure{%
  \@ifstar{\global\@dfloattrue \@bfigure}{\global\@dfloatfalse \@bfigure}%
}

\def\@bfigure#1{%
  \par
  \if@dfloat
    \ItemSPAN=\Double
    \TEMPDIMEN=\PageWidth
  \else
    \ItemSPAN=\Single
    \TEMPDIMEN=\ColumnWidth
  \fi
  \ifsinglecol
    \TEMPDIMEN=\PageWidth
  \else
    \ItemSTATUS=\InStack
    \ItemNUMBER=#1%
    \ItemTYPE=\Figure
  \fi
  \bgroup
    \hsize=\TEMPDIMEN
    \global\setbox\ItemBOX=\vbox\bgroup
      \eightpoint\nostb@ls{10pt}%
      \let\caption=\fig@caption
      \ifsinglecol \let\nofloat=\no@float\fi
}

\def\fig@caption#1{%
  \vskip 5.5pt plus 6pt%
  \bgroup 
    \eightpoint\nostb@ls{10pt}%
    \setbox\TEMPBOX=\hbox{#1}%
    \ifdim\wd\TEMPBOX>\TEMPDIMEN
      \noindent \unhbox\TEMPBOX\par
    \else
      \hbox to \hsize{\hfil\unhbox\TEMPBOX\hfil}%
    \fi
  \egroup
}

\def\endfigure{%
  \par\egroup 
  \egroup
  \ifsinglecol
    \if@here \midinsert\global\@herefalse\else \topinsert\fi
      \unvbox\ItemBOX
    \endinsert
  \else
    \JoinStack
    \Print{Processing source for figure \the\ItemNUMBER}%
  \fi
}


\newbox\tab@cap@box
\def\tab@caption#1{\global\setbox\tab@cap@box=\hbox{#1\par}}

\newtoks\tab@txt@toks
\long\def\tab@txt#1{\global\tab@txt@toks={#1}\global\table@txttrue}

\newif\iftable@txt  \table@txtfalse
\newif\if@dfloat    \@dfloatfalse

\def\begintable{%
  \@ifstar{\global\@dfloattrue \@btable}{\global\@dfloatfalse \@btable}%
}

\def\@btable#1{%
  \par
  \if@dfloat
    \ItemSPAN=\Double
    \TEMPDIMEN=\PageWidth
  \else
    \ItemSPAN=\Single
    \TEMPDIMEN=\ColumnWidth
  \fi
  \ifsinglecol
    \TEMPDIMEN=\PageWidth
  \else
    \ItemSTATUS=\InStack
    \ItemNUMBER=#1%
    \ItemTYPE=\Table
  \fi
  \bgroup
    \eightpoint\nostb@ls{10pt}%
    \global\setbox\ItemBOX=\vbox\bgroup
      \let\caption=\tab@caption
      \let\tabletext=\tab@txt
      \ifsinglecol \let\nofloat=\no@float\fi
}

\def\endtable{%
  \par\egroup 
  \egroup
  \setbox\TEMPBOX=\hbox to \TEMPDIMEN{%
    \eightpoint\nostb@ls{10pt}%
    \hss
    \vbox{%
      \hsize=\wd\ItemBOX
      \ifvoid\tab@cap@box
      \else
        \noindent\unhbox\tab@cap@box
        \vskip 5.5pt plus 6pt%
      \fi
      \box\ItemBOX
      \iftable@txt
        \vskip 10pt%
        \noindent\the\tab@txt@toks
        \global\table@txtfalse
      \fi
    }%
    \hss
  }%
  \ifsinglecol
    \if@here \midinsert\global\@herefalse\else \topinsert\fi
      \box\TEMPBOX
    \endinsert
  \else
    \global\setbox\ItemBOX=\box\TEMPBOX
    \JoinStack
    \Print{Processing source for table \the\ItemNUMBER}%
  \fi
}

\def\UnloadZoneA{%
\FirstZoneAtrue
 \Iteration=\z@
  \loop
   \ifnum\Iteration<\LengthOfStack
    \GetItemSTATUS{\Iteration}
    \ifnum\ItemSTATUS=\InZoneA
     \GetItemBOX{\Iteration}
     \ifFirstZoneA \vbox to \BodgeHeight{\vfil}%
     \FirstZoneAfalse\fi
     \unvbox\ItemBOX\ItemSep
     \LeaveStack{\Iteration}
     \else
     \advance\Iteration \@ne
   \fi
 \repeat
}

\def\UnloadZoneC{%
\Iteration=\z@
  \loop
   \ifnum\Iteration<\LengthOfStack
    \GetItemSTATUS{\Iteration}
    \ifnum\ItemSTATUS=\InZoneC
     \GetItemBOX{\Iteration}
     \ItemSep\unvbox\ItemBOX
     \LeaveStack{\Iteration}
     \else
     \advance\Iteration \@ne
   \fi
 \repeat
}


\def\ShowItem#1{
  {\GetItemAll{#1}
  \Print{\the#1:
  {TYPE=\ifnum\ItemTYPE=\Figure Figure\else Table\fi}
  {NUMBER=\the\ItemNUMBER}
  {SPAN=\ifnum\ItemSPAN=\Single Single\else Double\fi}
  {SIZE=\the\ItemSIZE}}}
}

\def\ShowStack{%
 \Print{}
 \Print{LengthOfStack = \the\LengthOfStack}
 \ifnum\LengthOfStack=\z@ \Print{Stack is empty}\fi
 \Iteration=\z@
 \loop
 \ifnum\Iteration<\LengthOfStack
  \ShowItem{\Iteration}
  \advance\Iteration \@ne
 \repeat
}

\def\B#1#2{%
\hbox{\vrule\kern-0.4pt\vbox to #2{%
\hrule width #1\vfill\hrule}\kern-0.4pt\vrule}
}


\newif\ifsinglecol   \singlecolfalse

\def\onecolumn{%
  \global\output={\singlecoloutput}%
  \global\hsize=\PageWidth
  \global\vsize=\PageHeight
  \global\ColumnWidth=\hsize
  \global\TextLeading=12pt
  \global\Leading=12
  \global\singlecoltrue
  \global\let\onecolumn=\relax
  \global\let\footnote=\sing@footnote
  \global\let\vfootnote=\sing@vfootnote
  \ninepoint 
  \message{(Single column)}%
}

\def\singlecoloutput{%
  \shipout\vbox{\PageHead\vbox to \PageHeight{\pagebody\vss}\PageFoot}%
  \advancepageno
  \ifplate@page
    \shipout\vbox{%
      \sp@pagetrue
      \def\sp@type{plate}%
      \global\plate@pagefalse
      \PageHead\vbox to \PageHeight{\unvbox\plt@box\vfil}\PageFoot%
    }%
    \message{[plate]}%
    \advancepageno
  \fi
  \ifnum\outputpenalty>-\@MM \else\dosupereject\fi%
}

\def\ItemSep{\vskip\ItemSepamount\relax}

\def\ItemSepbreak{\par\ifdim\lastskip<\ItemSepamount
  \removelastskip\penalty-200\ItemSep\fi%
}


\let\@@endinsert=\endinsert 

\def\endinsert{\egroup 
  \if@mid \dimen@\ht\z@ \advance\dimen@\dp\z@ \advance\dimen@12\p@
    \advance\dimen@\pagetotal \advance\dimen@-\pageshrink
    \ifdim\dimen@>\pagegoal\@midfalse\p@gefalse\fi\fi
  \if@mid \ItemSep\box\z@\ItemSepbreak
  \else\insert\topins{\penalty100 
    \splittopskip\z@skip
    \splitmaxdepth\maxdimen \floatingpenalty\z@
    \ifp@ge \dimen@\dp\z@
    \vbox to\vsize{\unvbox\z@\kern-\dimen@}
    \else \box\z@\nobreak\ItemSep\fi}\fi\endgroup%
}


\def\gobbleone#1{}
\def\gobbletwo#1#2{}
\let\footnote=\gobbletwo 
\let\vfootnote=\gobbleone

\def\sing@footnote#1{\let\@sf\empty 
  \ifhmode\edef\@sf{\spacefactor\the\spacefactor}\/\fi
  \hbox{$^{\hbox{\eightpoint #1}}$}\@sf\sing@vfootnote{#1}%
}

\def\sing@vfootnote#1{\insert\footins\bgroup\eightpoint\b@ls{9pt}%
  \interlinepenalty\interfootnotelinepenalty
  \splittopskip\ht\strutbox 
  \splitmaxdepth\dp\strutbox \floatingpenalty\@MM
  \leftskip\z@skip \rightskip\z@skip \spaceskip\z@skip \xspaceskip\z@skip
  \noindent $^{\scriptstyle\hbox{#1}}$\hskip 4pt%
    \footstrut\futurelet\next\fo@t%
}

\def\footnoterule{\kern-3\p@ \hrule height \z@ \kern 3\p@}

\skip\footins=19.5pt plus 12pt minus 1pt
\count\footins=1000
\dimen\footins=\maxdimen

\def\note#1#2{%
  \let\@sf=\empty \ifhmode\edef\@sf{\spacefactor\the\spacefactor}\/\fi
  #1\insert\footins\bgroup
    \eightpoint\b@ls{10pt}\rm
    \interlinepenalty\interfootnotelinepenalty
    \splitmaxdepth\dp\strutbox \floatingpenalty\@MM
    \leftskip\z@skip \rightskip\z@skip \spaceskip\z@skip \xspaceskip\z@skip
    \noindent\footstrut #1$\,$#2\strut\par
  \egroup
  \@sf\relax}


\def\landscape{%
  \global\TEMPDIMEN=\PageWidth
  \global\PageWidth=\PageHeight
  \global\PageHeight=\TEMPDIMEN
  \global\let\landscape=\relax
  \onecolumn
  \message{(landscape)}%
  \raggedbottom
}


\output{%
  \ifLeftCOL
    \global\setbox\LeftBOX=\vbox to \ZoneBSize{\box255\unvbox\ZoneBBOX
      \ifvoid\footins\else
        \vskip\skip\footins\unvbox\footins\fi
    }%
    \global\LeftCOLfalse
    \MakeRightCol
  \else
    \setbox\RightBOX=\vbox to \ZoneBSize{\box255\unvbox\ZoneBBOX
      \ifvoid\footins\else
        \vskip\skip\footins\unvbox\footins\fi
    }%
    \setbox\MidBOX=\hbox{\box\LeftBOX\hskip\ColumnGap\box\RightBOX}%
    \setbox\PageBOX=\vbox to \PageHeight{%
      \UnloadZoneA\box\MidBOX\UnloadZoneC}%
    \shipout\vbox{\PageHead\vbox to \PageHeight{\box\PageBOX\vss}\PageFoot}%
    \advancepageno
    \ifplate@page
      \shipout\vbox{%
        \sp@pagetrue
        \def\sp@type{plate}%
        \global\plate@pagefalse
        \PageHead\vbox to \PageHeight{\unvbox\plt@box\vfil}\PageFoot%
      }%
      \message{[plate]}%
      \advancepageno
    \fi
    \global\LeftCOLtrue
    \CleanStack
    \MakePage
  \fi
}


\Warn{\start@mess}

\newif\ifCUPmtplainloaded 
\ifprod@font
  \global\CUPmtplainloadedtrue
\fi

\def\mnmacrosloaded{} 

\catcode `\@=12 



\fi
\input epsf
\epsfverbosetrue
\pageoffset{-2.5pc}{0pc}

\onecolumn

\begintopmatter

\title{Radio Interferometers with wide bandwidths}

\author{Ravi Subrahmanyan}

\affiliation{Australia Telescope National Facility, CSIRO, 
	     Locked bag 194, Narrabri, NSW 2390, Australia}
\vskip 0.1 truecm
\shortauthor{R. Subrahmanyan}

\shorttitle{Radio Interferometers with wide bandwidths}

\abstract{
The Australia Telescope Compact Array and the Very Large Array are
currently being upgraded to operate with wide bandwidths; interferometers
dedicated to the measurement of cosmic microwave background anisotropies 
are being designed with large instantaneous bandwidths for high
sensitivity.  Interferometers with wide instantaneous bandwidths that do
not operate with correlators capable of decomposing the bands into narrow
channels suffer from `bandwidth smearing' effects in wide-field imaging.
The formalism of mosaic imaging is extended here to interferometers
with finite bandwidths to examine the consequences for the imaging 
of wide fields if very wide instantaneous bandwidths are used.  
The formalism presented here provides an understanding of the signal
processing associated with wide-band interferometers: mosaicing may 
be viewed as decomposing visibilities over wide observing bands and,
thereby, avoiding `bandwidth smearing' effects in wide-field imaging.
In particular, the formalism has implications for interferometer 
measurements of the angular power spectrum of cosmic microwave background
anisotropies: mosaic-mode observing with wide-band radio interferometers
decompose wide-band data and synthesize narrow filters in multipole 
space.  
}

\keywords{instrumentation: interferometers -- methods: 
observational -- techniques: interferometric -- cosmic microwave 
background -- radio continuum: general }

\maketitle

\section{Introduction}

When observing continuum radio sources,
the sensitivity of a radio interferometer depends on the square root of the
bandwidth, apart from other factors.  For this reason, arrays are usually 
designed to have the maximum continuum bandwidths permitted by the 
state of the art and, additionally, the instantaneous
observing bandwidths of existing interferometer arrays are often upgraded
over time as the technology associated with sampling rates, transmission 
bandwidths and correlator speeds improve.  

Interferometer arrays that
Fourier synthesize images of relatively bright celestial sources 
are often dynamic range limited owing to artefacts resulting from 
calibration errors and other systematics; it is in the imaging of
the relatively weak sources that thermal noise proves to be the limiting
factor.  The radio interferometer imaging of cosmic microwave background
(CMB) anisotropies, in total intensity and in polarization, is an example
of an astrophysical problem where the thermal noise limitation, arising
partly from the limited bandwidths in radio interferometers, provides a
strong argument in favour of using bolometers -- which are intrinsically
wide bandwidth devices -- for the detection instead of radio 
interferometers.

Interferometer bandwidths are limited by the state of the technology.
If the visibility correlations are to be measured by digital correlators,
the wide bandwidth analog signals from the antenna elements would
have to be sampled at appropriately high speeds.  Interferometers in use
today that have the widest bandwidths do not digitise the antenna
signals; their correlators are analog multipliers.  And an interesting
proposal for a dedicated CMB interferometer uses bolometer detectors
to measure the interferometer fringe (Ali et al.  2002).

When the celestial source that is being imaged has an intensity distribution
over a wide field, or sources distributed over a wide field are being 
imaged together, wide bandwidth interferometers have special problems and
the visibilities suffer from an effect known as `bandwidth smearing' (see,
for example, Cotton (1999) and Bridle (1999)).   An alternate
description of this effect is in Section 2 for the case where a 
simple two-element interferometer observes sources distributed 
over a wide field. In Section 3,
the formalism of wide-field mosaicing is extended
to a wide bandwidth interferometer.  The analysis shows that
just as mosaic imaging techniques may be adopted for overcoming
the limitations posed by the size of the antenna elements, mosaic imaging
may also be adopted for overcoming the limitations posed by the
size of the bandwidth.   Later sections discuss the use of wide bandwidth
interferometers in wide field surveys of the sky 
and measuring the angular power spectrum of CMB temperature anisotropies.

It may be noted here that the `bandwidth smearing' problem in wide-field 
Fourier synthesis imaging may be avoided by adopting `narrow'
bandwidth synthesis or `wide' bandwidth synthesis 
techniques (Cotton 1999). These methods may not only avoid the
problem but also improve the imaging fidelity.  However, these bandwidth
synthesis methods require that the interferometer 
correlations be measured separately in multiple frequency channels 
covering the observing band; in other words, a spectral-line correlator 
is required for the continuum imaging.  

\section{Wide field imaging}

Consider an idealised case where two isotropic antenna elements with 
infinitesimal effective area, operating at a wavelength $\lambda$ 
and corresponding frequency $\nu = c/\lambda$,
form an interferometer with spacing $b$; $c$ here denotes the speed
of light.  The propagation delays in
the receiver chains are assumed to have been equalised so that coherent 
radiation arriving in phase at the two antenna elements will produce a
zero-phase interferometer response.  The interferometer response,
to the brightness distribution over the sky, is the 
coherence function $C(s,\nu)$
at the spatial frequency $s$, where $s = b/\lambda$.  The sky brightness
distribution results in an EM field on the ground and 
the measured visibility 
represents the coherence between the EM fields at the locations of the
two antenna elements, this spatial coherence function depends only on the
spacing $b$ and its orientation with respect to the sky and not on the
absolute locations of the elements.  The coherence function 
may be expressed as an integral of the sky brightness 
$I(\vec\bmath{r}, \nu)$ over the celestial sphere:
$$
C(\vec\bmath{s},\nu) = \int I(\vec\bmath{r}, \nu) 
e^{-2 \pi i (\vec\bmath{s} \cdot \vec\bmath{r})} d\Omega, \eqno\stepeq
$$
where $\vec\bmath{s}$ is the baseline 
vector: $\vec\bmath{s} = \vec\bmath{b}/\lambda$, and $\vec\bmath{r}$ 
is a unit vector towards the sky solid angle 
element $d\Omega$ (Clark 1999).
It is assumed here that the radiation is spatially incoherent
across celestial sources and that the sources are extremely distant
as compared to the spacing $b$.

In the discussions that follow, we assume that the online integration
times are sufficiently small, so that `time-averaging smearing' effects
are negligible.  Additionally, we restrict the imaging to
sufficiently small sky regions so that  
a two-dimensional (2-D) Fourier transform relationship between the
sky brightness distribution and the spatial coherence function is 
a valid approximation.  Within this approximation, if
the unit vector $\vec\bmath{r_{\circ}}$ points towards the interferometer
phase centre, and the vector $\vec\bmath{\xi} = (\vec\bmath{r} - \vec\bmath{r_{\circ}})$
defines offsets from the phase centre, 
$\vec\bmath{s} \cdot \vec\bmath{r}$ in the kernal of equation~(1) may be
replaced by $\vec\bmath{s} \cdot \vec\bmath{\xi}$.

\subsection{Effects arising from the use of finite antenna apertures}

If the effective apertures of the antennas cover finite areas 
on the ground, the interferometer response will be an integral 
of the spatial coherence function over the spatial frequencies 
sampled by the baseline vectors between elements of one antenna
aperture and elements of the other aperture.  

To take a specific example, assume that the antennas forming 
the interferometer have identical
circular apertures, uniformly illuminated, with diameter $d$ and
with a spacing $b$ between their centres.  The antenna apertures
are assumed to be in the same plane as the interferometer baseline
vector and, therefore, the antenna pointing coincides with the
interferometer phase centre.  The antenna configuration on the 
ground (real space) is shown in panel~$a$ of Fig.~1.   The interferometer
response (measured visibility) is an integral of the spatial
coherence function over one of two circular regions, with radius $d/\lambda$, 
of the spatial frequency domain; these circular regions are located
symmetrically in the 2-D spatial frequency space with their
centres $b/\lambda$ from the origin.  These sampling regions are shown 
in panel~$b$ of Fig.~1.  The values of the coherence function 
at points in the spatial frequency plane that are inversion
symmetric about the origin are complex conjugates of each other
and the implementation of the complex correlator determines the
choice of the sampling region.  The spatial frequencies 
are sampled over the range $(b-d)/\lambda$ to $(b+d)/\lambda$ with a
weighting that linearly decreases from the centres of the circular
sampled regions to zero at the edges. For the special case where
the antenna apertures are adjacent and $b=d$, the coherence function
is sampled over the range 0--$2d/\lambda$.

Interferometers with finite antenna apertures provide visibilities
that are integrals over spatial frequency space: the weighted averaging
effectively results in a loss of information on the detailed
variations of the coherence function over distances in spatial
frequency that are smaller than $2d/\lambda$.  
The weighting function in spatial frequency
space has a Fourier transform that is the antenna far-field 
radiation power pattern and, therefore, 
in the sky domain, the averaging (in spatial frequency space) 
results in that the interferometer response
to brightness away from the antenna pointing centre is attenuated
by the primary beam pattern of the antennas.  The interferometer does
not respond to brightness outside the primary beam
and this is a problem for wide-field imaging.

To appreciate better the effect of using antennas with finite-sized
apertures, consider the case where 
a single discrete source is present at a location offset from the 
centre of the primary beam, which is also
the interferometer phase centre.  Let the vector 
$\vec\bmath{\xi} = (\vec\bmath{r} - \vec\bmath{r_{\circ}}$) define the source position.
The spatial coherence function at a location $\vec\bmath{s}$ in spatial
frequency domain will have a phase $-2 \pi (\vec\bmath{s} \cdot \vec\bmath{\xi})$.
Moving along the spatial frequency plane in a direction 
parallel to $\vec\bmath{\xi}$ (in a direction that has the same
direction cosines as $\vec\bmath{\xi}$), the phase will wind with period
$(1/\xi)$ wavelengths, where $\xi$ is the magnitude of the vector offset
$\vec\bmath{\xi}$.  The visibility measurements are a weighted 
average over circular
regions of diameter $2d/\lambda$, across which 
the phase winds $(2 d \xi / \lambda)$ times.  If the source is offset
by a distance $\xi = \lambda/(2d) = c/(2d\nu)$, the phase of the
coherence function winds through $2\pi$ radians across the sampled 
circle in spatial frequency domain and the response is severely attenuated.
It is this averaging of a rotating coherence function vector,
across spatial frequency space, that results in the loss of information
on the sky brightness distribution and is usually called the `primary-beam
attenuation'.  

\subsection{Bandwidth related effects}

A baseline vector $\vec\bmath{b}$, between an element of one of the apertures
forming the interferometer and an element of the other antenna aperture,
will sample the coherence function at the spatial 
frequency $\vec\bmath{b} \nu /c$.  Therefore, if the receivers and correlator
electronics operate with a finite bandwidth and provide an average visibility
measurement over a band covering the range of frequencies $\pm \Delta \nu$
around a centre frequency $\nu_{\circ}$, the two elemental apertures
with baseline $\vec\bmath{b}$ will average the coherence function over a trace
in 2-D spatial frequency space 
from $\vec\bmath{b} ( \nu_{\circ} - \Delta \nu)/c$ to
$\vec\bmath{b} ( \nu_{\circ} + \Delta \nu)/c$. 

The finite observing bandwidth results in that 
the measured visibility is an
average of the coherence function across spatial frequency space just as 
the finite apertures result in the visibilities being an average. 
This averaging across the finite band will result in a loss of information
on sky intensity distribution if the traces, corresponding to each pair of
elemental apertures, average over a varying coherence function.  

In the example we have been considering of an interferometer formed
between a pair of circular and uniformly illuminated apertures, the finite
bandwidth results in that the circular regions sampled in spatial
frequency space scale with increasing frequency.  The diameter of the
sampled regions scales from $2d( \nu_{\circ} - \Delta \nu)/c$ at the bottom
of the frequency range to $2d( \nu_{\circ} + \Delta \nu)/c$ at the highest 
observing frequency.  The centre of the circular region also shifts outwards
from $b( \nu_{\circ} - \Delta \nu)/c$ to 
$b( \nu_{\circ} + \Delta \nu)/c$.  The movement of the 
circular sampling region
in spatial frequency space, from one end of the observing band to
the other, is depicted in panel~$c$ of Fig.~1.

The averaging of the coherence function in the spatial 
frequency domain, owing to the finite antenna
size, is weighted by a function
that is the cross-correlation between the illumination
patterns of the two apertures that constitute the interferometer.
In the case of identical antennas, the weighting function is the
auto-correlation of the illumination, which is also the Fourier
transform of the far-field radiation power pattern.
The averaging that results from the finite bandwidth is a
separate averaging of the coherence function in the 
spatial frequency domain; in this case the weighting function is defined
by the bandpass response of the interferometer. The observed visibility
may, therefore, be written as a double integral of the coherence function 
in spatial frequency space:
$$
V(\vec\bmath{s_{\circ}}) =
\int \int \left\{ C(\vec\bmath{s},\nu) W_s(\vec\bmath{s}-\vec\bmath{s_{\circ}},\nu) 
d\vec\bmath{s} \right\} W_\nu(\nu-\nu_{\circ}) d\nu, \eqno\stepeq
$$
where $\vec\bmath{s_{\circ}}$ is the baseline
vector between the centres of the apertures, in wavelengths, at the centre
frequency $\nu_{\circ}$.  $W_s(\vec\bmath{s},\nu)$ is the auto-correlation 
of the aperture illumination at frequency $\nu$ and is, in general,
a frequency dependent sampling function in spatial frequency space.
$W_\nu(\nu)$ is the weighting function
corresponding to the bandpass shape.  The weighting functions are
assumed to be normalised so that their integrals are unity. 

Consider the coherence function at a baseline $\vec\bmath{b_{\circ}}$ owing
to a celestial source that is offset from the
interferometer phase centre and at a location defined by the vector
$\vec\bmath{\xi}$.  Additionally, let the source offset be along the
baseline; {\it i.e.}, let $\vec\bmath{s_{\circ}}$, which is equal to 
$\vec\bmath{b_{\circ}}/\lambda$,
and $\vec\bmath{\xi}$ be vectors with the same direction cosines in their
respective planes.  The coherence function at this spatial frequency
will have a phase $-2 \pi b_{\circ} \nu \xi / c$ and the phase will wind
through $2 b_{\circ} \xi \Delta \nu / c$ turns across the band of
$\pm \Delta \nu$.  A source offset by $c/(2 b_{\circ} \Delta \nu)$ will
have a coherence function that winds through $2 \pi$ radians across the
observing band.  It is the loss of information arising from this
averaging across the observing band that is called `bandwidth smearing'.

The number of phase turns across the observing
bandwidth will exceed the number of turns across the visibility
space covered by finite apertures if 
$(\Delta \nu / \nu_{\circ}) > (d / b_{\circ})$.
The averaging length, in spatial frequency space, 
owing to the finite aperture is proportional to $d / b_{\circ}$ whereas that
owing to the finite bandwidth depends on $\Delta \nu / \nu_{\circ}$. The
relative magnitudes of these two quantities decides which effect dominates
in any interferometer.

\section{Mosaicing with a wide band interferometer}

Scanning the sky with an interferometer, and using the scan
data to reconstruct the distribution in the coherence function across
the spatial frequencies sampled by the apertures forming the interferometer,
was proposed by Ekers \& Rots (1979).  
As far as I know, Rao \& Velusamy (1984)
were the first to explicitely implement this observing scheme and decompose 
interferometer visibility data in the spatial frequency domain 
and then use the finely sampled visibility data to reconstruct
wide field images. I extend this formalism 
below to the case where an interferometer
mosaic observes a wide field using a very wide bandwidth. 

Consider the case where the arms of a two-element interferometer 
are phased towards the phase centre $\vec\bmath{r_{\circ}}$ and the individual
antennas also point towards $\vec\bmath{r_{\circ}}$ so that the offset in the 
pointing of the antennas from the phase centre is zero.
The observed visibility in this case is 
$$
V(\vec\bmath{s_{\circ}}, 0) =
\int \int \left\{ C(\vec\bmath{s},\nu) W_s(\vec\bmath{s}-\vec\bmath{s_{\circ}},\nu) 
d\vec\bmath{s} \right\} W_\nu(\nu-\nu_{\circ}) d\nu. \eqno\stepeq
$$

When the pointing of the antennas is changed from $\vec\bmath{r_{\circ}}$
to $\vec\bmath{r}$, with the offset denoted by
$\vec\bmath{\xi} = (\vec\bmath{r}-\vec\bmath{r_{\circ}})$,
the coherence function measured between pairs of elements
on the two apertures now picks up an additional phase gradient.  
If the antennas are co-mounted on a common platform,
and the pointing as well as phase centres are offset, together, 
corresponding to the vector $\vec\bmath{\xi}$, the additional phase is given by 
$$
\phi(\vec\bmath{s}) = 2 \pi \left\{ \vec\bmath{\xi} \cdot \vec\bmath{s} 
\right\}. \eqno\stepeq
$$
If the antennas are independently mounted, and the 
interferometer phase centre
is kept unchanged, the additional phase acquired at
any spatial frequency $\vec\bmath{s}$ is 
$$
\phi(\vec\bmath{s}) = 2 \pi \left\{ \vec\bmath{\xi} \cdot (\vec\bmath{s} 
- \vec\bmath{b_{\circ}} \nu / c)\right\}. \eqno\stepeq
$$
In this second case, the signals from the individual antennas may be delayed
to introduce a differential delay 
of $\tau = (\vec\bmath{\xi} \cdot \vec\bmath{b_{\circ}})/c$ between the antenna pair
with baseline $\vec\bmath{b_{\circ}}$.  Then the pointing, delay and phase centres of
the independently mounted array would be offset through $\vec\bmath{\xi}$ and
the additional phase would be given by equation~(4). Assuming that such
differential delays are introduced in the case of arrays with independently
mounted antennas, the observed visibility is given by (for both mounts)
$$
\eqalignno{%
V(\vec\bmath{s_{\circ}}, \vec\bmath{\xi}) &	=
\int \int \left\{ C(\vec\bmath{s},\nu) W_s(\vec\bmath{s}-\vec\bmath{s_{\circ}},\nu) 
e^{i 2 \pi ( \vec\bmath{\xi} \cdot \vec\bmath{s} )}
d\vec\bmath{s} \right\} W_\nu(\nu-\nu_{\circ}) d\nu. & \startsubeq\cr
 & = \int \int \left\{ 
C(\vec\bmath{s},\nu) W_s(\vec\bmath{s}-\vec\bmath{s_{\circ}},\nu) 
W_\nu(\nu-\nu_{\circ}) \right\}
e^{i 2 \pi ( \vec\bmath{\xi} \cdot \vec\bmath{s} )}
d\vec\bmath{s}  d\nu. & \stepsubeq\cr}
$$

Consider the case where the two-element interferometer scans the sky, 
moving the pointing of the antennas as well as the phase and delay
centres together over a range in offset $\vec\bmath{\xi}$.  This is achieved
in the co-mounted case by simply tipping the platform over a range of angles.
If the visibilities $V(\vec\bmath{s_{\circ}}, \vec\bmath{\xi})$ are accumulated
over a range of $\vec\bmath{\xi}$ and Fourier transformed to give
$$
V(\vec\bmath{s_{\circ}}, \vec\bmath{\chi}) = \int V(\vec\bmath{s_{\circ}}, \vec\bmath{\xi})
e^{-i 2 \pi ( \vec\bmath{\xi} \cdot \vec\bmath{\chi} )} d\vec\bmath{\xi}, \eqno\stepeq
$$
we obtain 
$$
\eqalignno{%
V(\vec\bmath{s_{\circ}}, \vec\bmath{\chi}) & = 
\int \int \int \left\{  C(\vec\bmath{s},\nu) W_s(\vec\bmath{s}-\vec\bmath{s_{\circ}},\nu) 
W_\nu(\nu-\nu_{\circ}) \right\}
e^{i 2 \pi [ \vec\bmath{\xi} \cdot ( \vec\bmath{s} - \vec\bmath{\chi} ) ] }
d\vec\bmath{s}  d\nu d\vec\bmath{\xi} & \startsubeq\cr
& = \int \left\{ \int C(\vec\bmath{s},\nu) W_s(\vec\bmath{s}-\vec\bmath{s_{\circ}},\nu) 
W_\nu(\nu-\nu_{\circ}) d\nu \right\}
\delta(\vec\bmath{s} - \vec\bmath{\chi}) d\vec\bmath{s} & \stepsubeq\cr
& = \left\{ \int C(\vec\bmath{s},\nu) W_s(\vec\bmath{s}-\vec\bmath{s_{\circ}},\nu) 
W_\nu(\nu-\nu_{\circ}) d\nu \right\} 
\mid _{\rm{evaluated~at~}\vec\bmath{s}=\vec\bmath{\chi}} & \stepsubeq\cr
& = \int C(\vec\bmath{\chi},\nu) W_s(\vec\bmath{\chi}-\vec\bmath{s_{\circ}},\nu) 
W_\nu(\nu-\nu_{\circ}) d\nu. & \stepsubeq\cr}
$$
The $\delta$-function above represents the Dirac $\delta$-function.
The Fourier transformation of the visibility data acquired in different
pointings is seen to yield a weighted average of the coherence
function.  The averaging in equation~(8d) is over frequency and not over spatial
frequency and, therefore, this method of deriving the 
visibilities in spatial frequency domain is capable of avoiding the problems
associated with band-width smearing related effects.

If the coherence function $C(\vec\bmath{\chi},\nu)$ is independent of
frequency and only a function of the spatial frequency $\vec\bmath{\chi}$,
$$
V(\vec\bmath{s_{\circ}}, \vec\bmath{\chi}) =
C(\vec\bmath{\chi}) \int W_s(\vec\bmath{\chi}-\vec\bmath{s_{\circ}},\nu) 
W_\nu(\nu-\nu_{\circ}) d\nu. \eqno\stepeq
$$ 
In this case, the Fourier transformation of the visibilities that were acquired
with the interferometer pointed at different offsets 
yields weighted samples of the coherence function;
the relative weighting depends on the aperture illuminations 
of the antennas forming the interferometer pair and the bandpass shape.

\section{Wide field imaging}

In practice, the visibilities $V(\vec\bmath{s_{\circ}}, \vec\bmath{\xi})$ might 
be accumulated in a 2-D discrete grid of sky angle offsets spanning a range
$\pm \xi_m$ and with grid size $\Delta \xi$.  
Let this `bed-of-nails' sampling function (a comb of Dirac $\delta$-functions) 
in the sky domain be
denoted by $\Xi(\vec\bmath{\xi})$. For the case where
apertures of diameter $d$ have their centres spaced $b$ apart, and 
the observing band is over the range $\pm \Delta \nu$, each complex 
visibility measurement is composed of spatial frequencies 
in the range from  $(b-d)(\nu_{\circ} - \Delta\nu)/c$
to $(b+d)(\nu_{\circ} + \Delta\nu)/c$.  The complex visibilities (of the
real sky) are band limited in the spatial frequency domain to a range of
$$
\Delta s = { {2(d \nu_{\circ} + b \Delta \nu)} \over c} 
= 2 \left( d + b {{\Delta \nu}\over{\nu_{\circ}}} \right) 
{ {\nu_{\circ}} \over c } {\rm~~wavelengths}, \eqno\stepeq
$$
which equals approximately $2 d \nu_{\circ} / c$ 
when the bandwidth is small and $(\Delta \nu / \nu_{\circ}) \ll (d/b)$.
Nyquist sampling, of this signal that is complex and band limited in the 
spatial frequency domain, requires that 
$$
\Delta \xi < {c \over {2(d \nu_{\circ} + b \Delta \nu)} }. \eqno\stepeq
$$
It may be noted here that if the mosaic observations are made using an
interferometer array of antennas, all with diameters $d$ and observing
over a fixed band covering a range $\pm \Delta \nu$ around a centre frequency
$\nu_{\circ}$, avoiding aliasing in the spatial frequency domain requires
that equation~(11) be satisfied for the longest baseline. 

Explicit transformation of the visibilities $V(\vec\bmath{s_{\circ}},\vec\bmath{\xi})$
into visibilities $V(\vec\bmath{s_{\circ}}, \vec\bmath{\chi})$ distributed
over the spatial frequency range given by equation~(10) is possible
using equation~(9) provided that the Nyquist sampling criterion is satisfied.
The discrete Fourier transformation (DFT) of the visibilities 
$V(\vec\bmath{s_{\circ}}, \vec\bmath{\xi})$, which are discrete samples in
offset $\vec\bmath{\xi}$ space, would yield samples of the visibility  
$V(\vec\bmath{s_{\circ}}, \vec\bmath{\chi})$ in spatial frequency.  
If the mosaic grid measures $2 \xi_m / \Delta \xi$ complex visibilities  
along any sky dimension, the DFT would yield   
$2 \xi_m / \Delta \xi$ independent measures of the 
complex visibility in the conjugate space which is the
spatial frequency $\vec\bmath{\chi}$ domain.  Samples
spaced $1/(2 \xi_m)$ wavelengths apart would be independent. These
visibility samples are weighted averages of the coherence function
in spatial frequency domain with a point spread function (PSF)
that is determined by the sampling in $\vec\bmath{\xi}$ space
and is independent of the bandwidth; the samples 
$V(\vec\bmath{s_{\circ}}, \vec\bmath{\chi})$ are essentially averages 
of the coherence function over spatial frequency ranges that 
are $1/(2 \xi_m)$ wavelengths wide.  

The implication here is that even if the observing bandwidth is large,
and in the extreme case if $(\Delta \nu / \nu_{\circ})$ exceeds 
$(d/b)$ so that the attenuation owing to bandwidth smearing exceeds
the primary beam attenuation in the individual fields, mosaicing
observations with a sampling that satisfies equation~(11) can
image large fields of view.  The DFT to spatial frequencies would
decompose the observed visibilities -- that are integrals of the
coherence function over wide spatial frequency ranges
corresponding to the wide bandwidths used -- into samples that correspond to
regions that are effectively $1/(2 \xi_m)$ wavelengths wide. 
{\it Just as mosaicing has the ability to decompose the observations made
with wide-aperture antennas into visibilities corresponding to those
obtained with small aperture arrays, mosaic mode observing has the
ability to decompose the visibilities made with wide bandwidths into
those corresponding to narrow bands. }

The DFT to spatial frequencies
decomposes the observed visibilities into $1/(2 \xi_m)$ wavelength wide
bins and, as stated above, these visibilities 
$V(\vec\bmath{s_{\circ}}, \vec\bmath{\chi})$ are samples of the 
coherence function, weighted by functions that are defined by the aperture
illumination and bandpass shape, and convolved by a PSF 
$\Xi^{\prime}(\vec\bmath{\chi})$ that is the Fourier transform of the sampling
function $\Xi(\vec\bmath{\xi})$:
$$
V(\vec\bmath{s_{\circ}}, \vec\bmath{\chi}) =
\left\{ C(\vec\bmath{\chi}) \int W_s(\vec\bmath{\chi}-\vec\bmath{s_{\circ}},\nu) 
W_\nu(\nu-\nu_{\circ}) d\nu \right\} 
\otimes \Xi^{\prime}(\vec\bmath{\chi}), \eqno\stepeq
$$
where the operator $\otimes$ denotes convolution. The weighting function
$$
W_{s\nu}(\vec\bmath{\chi}) = \int W_s(\vec\bmath{\chi}-\vec\bmath{s_{\circ}},\nu) 
W_\nu(\nu-\nu_{\circ}) d\nu \eqno\stepeq
$$
is computable from the aperture illuminations of the antennas forming
the interferometer pair and the bandpass shape, $\Xi^{\prime}(\vec\bmath{\chi})$
is also computable from the adopted mosaic grid on the sky.  In terms of
these known functions, the derived visibilities may be 
expressed in the form
$$
V(\vec\bmath{s_{\circ}}, \vec\bmath{\chi}) = 
\left\{ C(\vec\bmath{\chi}) W_{s\nu}(\vec\bmath{\chi}) \right\}
\otimes \Xi^{\prime}(\vec\bmath{\chi}). \eqno\stepeq
$$
Samples of these visibility estimates, which are separated by 
$1/(2 \xi_m)$ wavelengths, are independent and may be used
to reconstruct images of the sky.
These  images are that of the true sky intensity distribution convolved by
a PSF that is the Fourier transform of the weights $W_{s\nu}(\vec\bmath{\chi})$.
The image would be tapered by the DFT of $\Xi^{\prime}(\vec\bmath{\chi})$, which
is simply the top-hat function $\Xi(\vec\bmath{\xi})$ that encompasses the entire
sky area covered by the mosaic pattern. 

\section{The power spectrum of CMB anisotropies}

Assuming that the sky temperature anisotropies in the CMB are Gaussian
random fluctuations with random phase, they may be completely described
by the $C_{l}$ coefficients of spherical harmonic decompositions.  
Observations of CMB anisotropies attempt to measure the angular power spectrum,
which is the distribution of $l(l+1)C_{l}/(2 \pi)$ over multipole $l$ space.

The $C_{l}$ coefficient represents the anisotropy power at multipole mode $l$
and in the limit of large $l$ and small sky angles, the spherical harmonic
decomposition approximates to a Fourier decomposition and
the distribution of anisotropy power over $l$ space is simply related 
to the distribution of the variance in observed visibilities over 
spatial frequencies.  A spatial frequency $s_{\circ}$ corresponds 
to a multipole mode $l_{\circ} = 2 \pi s_{\circ}$ and the variance 
in CMB visibilities that are measured over a band $\Delta s$ is a measure
of the CMB anisotropy over a multipole range $\Delta l = 2 \pi \Delta s$.
Interferometer measurements of CMB anisotropy hence provide estimates
of $C_l$ power in which the telescope filter functions are 
defined by the aperture illuminations of the antennas and the 
projected baseline length.
If the interferometer mosaic images a wide field, these visibility data 
at multiple pointings might be used to derive visibilities corresponding 
to narrow filters in spatial frequency space.  Therefore, mosaic mode
observations are a method for decomposing the interferometer $C_{l}$
measurements into those corresponding to narrow telescope filter 
functions in $l$ space: mosaic
observing improves $l$-space resolution (White et al. 1999; Subrahmanyan 2002).
Moreover, it has been argued that drift scanning the sky, using a co-mounted
antenna array operated as interferometers, is a useful technique for
rejecting systematics (Subrahmanyan 2002); drift scanning is an implementation of
the mosaicing technique. 

Towards any single pointing, a wide bandwidth interferometer samples 
a range in $l$ space and the anisotropy variance is averaged over this
range to produce the interferometer response.
When a wide-band interferometer observes CMB anisotropies over a wide
field in mosaic mode, as discussed in the preceeding sections, 
the visibilities obtained
in the multiple pointings may be transformed to yield visibilities
distributed in spatial frequency $\chi$.  In the small angle
approximation where $\vec\bmath{l} = 2 \pi \vec\bmath{\chi}$, 
$$
V(\vec\bmath{s_{\circ}}, \vec\bmath{l}) = \left\{ 
C(\vec\bmath{l}) W_{s\nu}(\vec\bmath{l}) \right\}
\otimes \Xi^{\prime}(\vec\bmath{l}). \eqno\stepeq
$$
The variance in $V(\vec\bmath{s_{\circ}}, \vec\bmath{l})$ is an estimate of the
anisotropy power at multipole order $l = \mid \vec\bmath{l} \mid$ and this is
independent of the instantaneous bandwidth of the interferometer.
The mosaic mode effectively decomposes the measurements that are averages over
wide bandwidths -- and, consequently, averages over wide $l$-space domains --
into narrow $l$-space filters that are defined by $\Xi^{\prime}(\vec\bmath{l})$.

\section{Sensitivity considerations}

Consider the case where a two-element interferometer, consisting
of apertures of diameter $d$ and baseline $b$, observes an 
$n \times n$ mosaic of pointings for a total time $t$ (the time spent 
at each pointing is $t/n^{2}$).  Assume first that the observations
are made using a narrow bandwidth $(\Delta \nu / \nu_{\circ}) \ll (d/nb)$.
Assume further that the pointings are separated by the required Nyquist rate
of $(c/2d\nu_{\circ})$.  The signal to noise ratio 
(SNR) in the measurement of the 
flux density of a point source that appears in any one of the 
pointings will be proportional to $d^{2} \sqrt{t} / n$.  If, instead, 
this interferometer that is formed of apertures of diameter $d$
observes the source position for the entire time $t$, the SNR would be
proportional to $d^{2} \sqrt{t}$.  

If the interferometer consisted of a pair of
apertures of size $d/n$ and the entire field were observed as a single
pointing for the same total time $t$, the SNR would be proportional to
$d^{2} \sqrt{t} / n^{2}$.  Using $n^{2}$ such apertures with diameter $d/n$
and simultaneously measuring $n^{2}$ correlations recovers the SNR that is proportional
to $d^{2} \sqrt{t} / n$.  If an array consisting of 
$2 \times n^{2}$ such apertures were used, with a total collecting area
corresponding to the area of the two-element interferometer with 
apertures of diameter $d$, and if all 
$n^{4}$ correlations were simultaneously measured, the SNR would
increase further and be proportional to $d^{2} \sqrt{t}$. 
This last value is the same as that expected in the case
where a two-element interferometer with apertures of diameter $d$ 
observes a source for the entire time $t$; however, the observation 
with apertures of size $d/n$ provides this high sensitivity over the entire mosaic
field which is larger by factor $n \times n$.

If the fractional bandwidth exceeds $d/(nb)$ -- so that bandwidth smearing 
effects are significant over the wide mosaiced field -- but if the
constraint $(\Delta \nu / \nu_{\circ}) \ll (d/b)$ is satisfied, mosaic
observations of the wide field using the $d$-sized apertures and covering
the wide field in $n \times n$ pointings would not
be compromised in sensitivity.  However, non-mosaiced observations of
the wide field using $d/n$-sized apertures and the wide bandwidths would 
suffer from  bandwidth smearing effects.

If $(\Delta \nu / \nu_{\circ}) \ga (d/b)$, the number of mosaic pointings
would have to be significantly increased.  
If $(\Delta \nu / \nu_{\circ}) = k (d/b)$,
the pointing grid size $\Delta \xi$ would have to be reduced by 
factor $(1+k)$.  A mosaic observation using a two-element interferometer
consisting of apertures of diameter $d$, covering a
grid of size $n(1+k) \times n(1+k)$, would detect point sources in the
wide field with a
$$
{\rm SNR} \propto \left( {{d^{2} \sqrt{t}}\over{n}} \right) 
{{\sqrt{k}} \over {(1+k)}}. \eqno\stepeq
$$
This SNR has a maximum at $k=1$ corresponding to the case where 
$(\Delta \nu / \nu_{\circ}) = (d/b)$ and the mosaic is made with a grid
size on the sky of 
$$
\Delta \xi = {c \over {4d \nu_{\circ}} }. \eqno\stepeq
$$

The analysis suggests that when interferometers with wide instantaneous bandwidths 
are used for the  mosaic imaging of wide fields, including the particular case of 
mosaic-mode CMB observations, the optimum bandwidth corresponds to the case where 
$(\Delta \nu / \nu_{\circ}) = (d/b)$; the SNR degrades if the instantaneous bandwidth
used exceeds $(2d/b)\nu_{\circ}$. 

Interferometers with wide fractional bandwidths are of particular interest at high
frequencies where errors due to telescope pointing and interferometer phase 
may be significant. therefore, the limitations arising from such errors are
considered below.

Any systematic pointing error of magnitude $\epsilon_{p}$ 
in the antennas forming the wide-bandwidth interferometer
would result in phase errors in the complex visibilities 
following the decomposition in spatial frequency space. These errors
would vary across spatial frequencies and lie in the range 
($\epsilon_{p}/\theta_{\nu_{\circ} - \Delta \nu}$) to 
($\epsilon_{p}/\theta_{\nu_{\circ} + \Delta \nu}$), where 
$\theta_{\nu}$ is the FWHM of the primary beam at frequency $\nu$.
The limiting factors to the quality of mosaic images made using interferometer
arrays was discussed by Cornwell, Holdaway \& Uson (1993) for the case where the
visibility was obtained in narrow bands.  In the case of mosaic imaging with a 
wide bandwidth interferometer, the image fidelity (defined as the ratio of the
value of an image pixel to the error between the true sky and reconstructed image) is,
in particular, limited by telescope pointing errors.  The fidelity is limited
approximately to
$$
\Lambda^{PE} \approx { {\sqrt{N_{A}} \theta_{\nu_{\circ}} \over 
{\epsilon_{p} \left( 1 + {{b \Delta \nu}\over{d \nu_{\circ}}} \right)}} }, \eqno\stepeq
$$
where $\theta_{\nu_{\circ}}$ is the FWHM of the primary beam 
at frequency $\nu_{\circ}$, $N_{A}$ is
the number of antenna elements in the array and $\epsilon_{p}$
is the systematic pointing error associated with any antenna; it is 
assumed that the errors are uncorrelated between antennas.

Mosaic scanning is also limited by time varying phase errors that introduce
a varying phase error over the sky scans.  If $\epsilon_{\phi}$ is the R.M.S.
phase error (in radians) during the scanning then the visibility amplitudes 
following the decomposition in spatial frequency space would have fractional 
errors of order $\epsilon_{\phi}$.  In the case of interferometer mosaic imaging,
assuming that these errors are antenna
based (as would be expected for atmospheric phase errors), the image plane
fidelity would be approximately limited by these errors to $N_{A}/\epsilon_{\phi}$.

\section{Summary}

Wide band interferometers -- in which the band is not
finely sub-divided in multi-channel receivers -- instantaneously sample
a wide domain in spatial frequency space.  Mosaic imaging of wide fields,
which are made by covering the wide fields with a grid of pointings 
and subsequently transforming the visibilities that are measured in 
the multiple pointings to visibility space, yields samples of the 
visibilities distributed in the spatial frequency domain.  
The formalism of mosaic imaging of wide fields has been extended
here to the case where the interferometers operate with wide
instantaneous bandwidths.  The mosaicing technique may be viewed as 
effectively decomposing wide-band data into spatial frequency bins.
The formalism presented here develops the understanding of the 
mosaicing technique that reconstructs wide-field images without 
`bandwidth smearing' effects. 

The signal-to-noise ratio in a mosaicing interferometer improves as the 
bandwidth is increased; however, the usefulness of the wide band for wide-field
imaging diminishes once the bandwidth exceeds $(2d/b)\nu_{\circ}$.
Admittedly, a multi-channel receiver improves upon the sensitivity
of the mosaicing interferometer; however, the gain is marginal if the
total band is less than $(2d/b)\nu_{\circ}$. The considerations discussed
here are relevant both to the case where the mosaicing interferometer
attempts to reconstruct sky images and to the case where the mosaicing
interferometer attempts a measurement in spatial frequency space of the 
angular power spectrum of the sky brightness fluctuations.

\section*{Acknowledgments}

This work was stimulated by discussions with members of the Taiwanese AMiBA
CMB project.

\beginfigure*{1}
\epsfbox{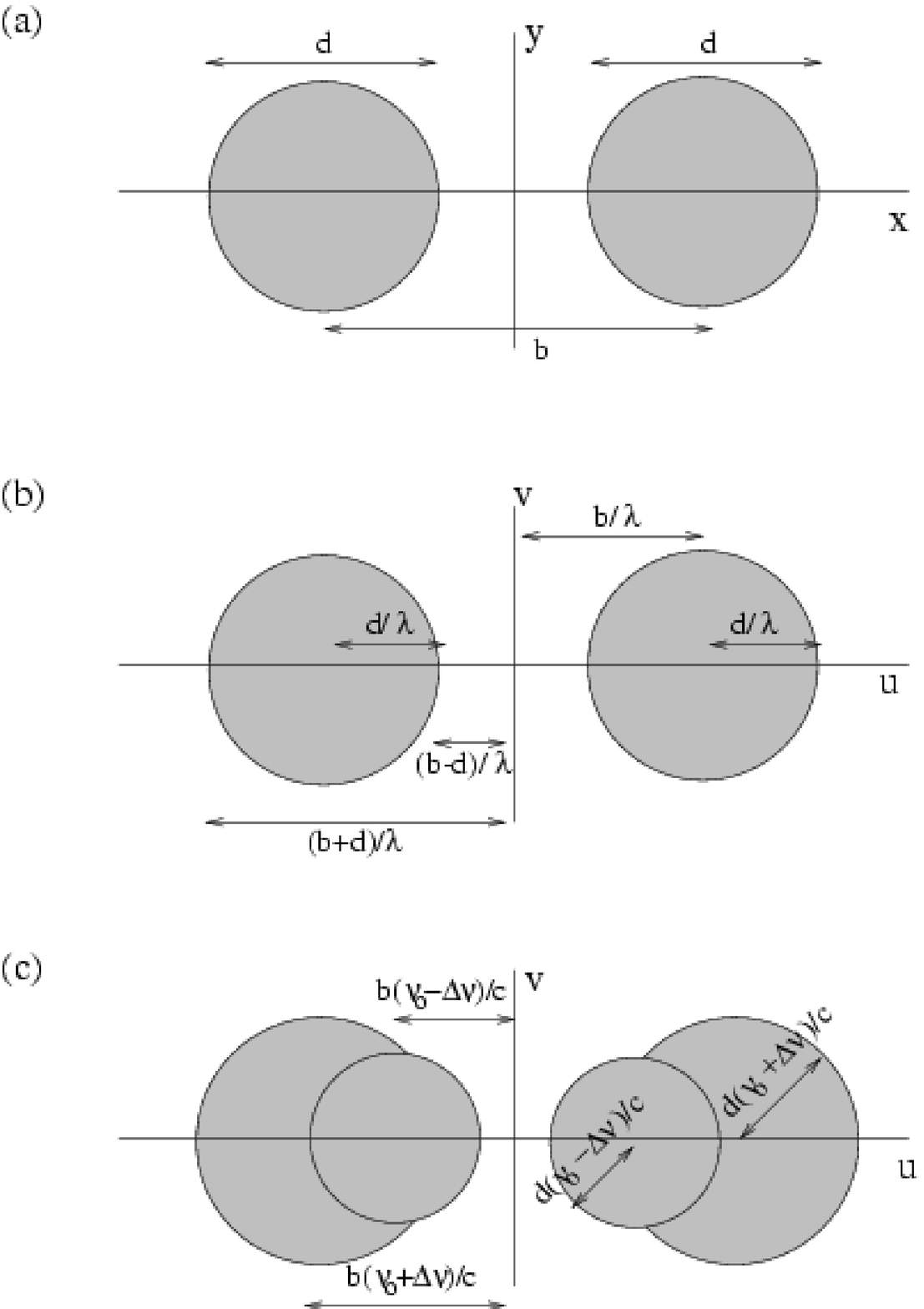}
\caption{{\bf Figure~1.} 
Panel $a$ shows the configuration and dimensions of the aperture
elements forming the interferometer in real space.  Panel $b$ shows the
visibility sampling in spatial frequency space. Panel $c$ shows the change
in the sampling of the spatial frequency domain across the observing
band.
 }
\endfigure

\end